\documentclass[a4paper, 11pt]{article}
\usepackage[a4paper, margin=1in]{geometry}  
\usepackage{setspace}  
\usepackage{graphicx}  
\usepackage{amsmath, amssymb}  
\usepackage{booktabs}  
\usepackage{array}  
\usepackage{times}  
\usepackage{amsmath, amssymb}  
\usepackage{cancel}  
\usepackage{graphicx}
\usepackage{slashed}
\usepackage{hyperref}  
\usepackage{listings}  
\usepackage{color}     
\usepackage{cite}      
\usepackage{float}
\usepackage{array}
\usepackage[xcdraw]{xcolor}
\usepackage{multirow}
\providecommand{\openone}{\leavevmode\hbox{\small1\kern-3.8pt\normalsize1}}

\usepackage{anyfontsize}

\usepackage[T1]{fontenc}

\usepackage{courier}  
\usepackage{adjustbox}
\begin{document}
\thispagestyle{empty}
\hfill IFJPAN-IV-2026-10

\def\thefootnote{\fnsymbol{footnote}}

\begin{center}
\Large\boldmath\textbf{
Impact of hidden heavy Higgs channels of VLB-Quarks below 1\,TeV in 2HDM
   }
\unboldmath
\end{center}
\vspace{-0.5cm}
\begin{center}
R. Benbrik$^a$  \footnote{
   	\href{mailto:r.benbrik@uca.ac.ma}{r.benbrik@uca.ac.ma} 	
}, M. Berrouj $^a$ \footnote{ \href{mailto:mbark.berrouj@ced.uca.ma}{mbark.berrouj@ced.uca.ma}}, M. Boukidi$^b$ \footnote{ \href{mailto:mohammed.boukidi@ifj.edu.pl}{mohammed.boukidi@ifj.edu.pl}}, M. Ech-chaouy$^a$  \footnote{
\href{mailto:m.echchaouy.ced@uca.ac.ma}{m.echchaouy.ced@uca.ac.ma}},
K. Kahime $^c$ \footnote{ \href{mailto:Kahimek@gmail.com}{Kahimek@gmail.com}}
, K. Salime$^a$  \footnote{
\href{mailto:k.salime.ced@uca.ac.ma}{k.salime.ced@uca.ac.ma}}

\textsl{\small
    $^a$Polydisciplinary Faculty, Laboratory of Physics, Energy, Environment, and Applications, Cadi Ayyad University, Sidi Bouzid, B.P. 4162, Safi, Morocco.\\
    $^b$Institute of Nuclear Physics, Polish Academy of Sciences, ul. Radzikowskiego 152, Cracow, 31-342, Poland.\\$^c$Laboratoire Interdisciplinaire de Recherche en Environnement, Management, Energie et Tourisme (LIREMET), ESTE, Cadi Ayyad University, B.P. 383, Essaouira, Morocco.\\
    }

\end{center}
\vspace{0.2cm}

\renewcommand{\thefootnote}{\arabic{footnote}}
\setcounter{footnote}{0}

\begin{abstract}
We investigate the phenomenological impact of incorporating vector-like bottom (VLB) quarks into the Type-II Two-Higgs-Doublet Model (2HDM-II). This framework introduces novel beyond-Standard-Model (BSM) decay channels $B \to Hb$, $B \to Ab$, and $B \to H^-t$, which are typically ignored by LHC pair-production searches focused on Standard Model (SM) final states ($B \to Zb$, $B \to hb$, $B \to Wt$). Our analysis reveals that these BSM pathways significantly weaken current VLB mass constraints. In the 2HDM-II alignment limit, the mass limit for a singlet $B$ shifts from approximately 1.5 TeV down to 1.34 TeV. For $(T, B)$ and $(B, Y)$ doublet configurations, the mass limits relax further to approximately 0.98 TeV, driven by the dominance of $B \to Hb$ and $B \to Ab$ decays, which can reach combined branching ratios of nearly 100\%.

\end{abstract}

\clearpage

\tableofcontents

\clearpage

\section{Introduction}
The discovery of a Higgs boson with a mass around 125~GeV at the LHC~\cite{ATLAS:2012yve, CMS:2012qbp} confirmed the Standard Model (SM) as a successful low-energy theory of electroweak interactions. However, the scalar sector may not be minimal. The Two-Higgs-Doublet Model (2HDM)~\cite{Gunion:1992hs, Branco:2011iw} provides a well-motivated extension, predicting additional Higgs bosons: a heavy CP-even scalar ($H$), a CP-odd pseudoscalar ($A$), and a charged Higgs pair ($H^\pm$). These states are actively searched for at the LHC but remain elusive.

The Vector-like quarks (VLQs)~\cite{Aguilar-Saavedra:2009xmz, Okada:2012gy, Buchkremer:2013bha, Han:2025itd, Han:2022jcp, Han:2022zgw, Yang:2025ktj, Wang:2013jwa, He:1999vp, He:2014ora, Yang:2024aav} are hypothetical fermions whose left- and right-handed components transform identically under the SM electroweak gauge group 
$SU(2)_L \otimes U(1)_Y$, in contrast to the chiral nature of the SM quarks. 
They naturally emerge in a variety of BSM frameworks, including models with extra dimensions~\cite{Chang:1999nh, Gherghetta:2000qt, Contino:2003ve}, Little Higgs \cite{Arkani-Hamed:2002iiv, Schmaltz:2002wx, Chang:2003vs, Han:2003wu}, composite Higgs models~\cite{Agashe:2004rs, Bellazzini:2014yua, Contino:2006qr, Lodone:2008yy, Matsedonskyi:2012ym, He:2001fz}, and grand unified theories~\cite{Hewett:1988xc} offer rich collider signatures. These color-triplet, spin-$1/2$ fermions can acquire vector-like mass terms that are independent of electroweak symmetry breaking and are typically organized into singlets ($T$, $B$), doublets [($T$, $B$), ($X$, $T$), ($Y$, $B$)], and triplets. At hadron colliders, VLQs are predominantly pair-produced via QCD interactions. Consequently, the production cross section depends only on the VLQ mass and the collider center-of-mass energy. 

Current LHC searches have predominantly targeted VLQs decaying into SM bosons ($W$, $Z$, and $h$), resulting in stringent lower bounds on their masses. For example, a vector-like $T$ quark decaying exclusively into $Wb$ is excluded up to $\sim 1.7~\text{TeV}$~\cite{ATLAS:2024gyc}, while the exclusive decay $B \to hb$ is constrained up to $\sim 1.58~\text{TeV}$~\cite{CMS:2024bni}. When non-exclusive decay patterns are considered, the limits become representation-dependent: singlet and doublet $T$ quarks are excluded up to approximately $1.49~\text{TeV}$ and $1.5~\text{TeV}$, respectively~\cite{CMS:2022fck}, whereas singlet and doublet $B$ quarks are constrained up to about $1.49~\text{TeV}$ and $1.52~\text{TeV}$~\cite{CMS:2024bni}. These bounds, however, rely on the implicit assumption that VLQs decay exclusively into SM final states.

When embedded in extended scalar sectors such as the Type-II 2HDM, VLQs can also decay into non-standard Higgs bosons, including $H^\pm$, $H$, and $A$. They can dominate in specific regions of parameter space and significantly alter collider sensitivities. Previous studies have examined the phenomenology of VLQs within the 2HDM Type-II framework~\cite{Gopalakrishna:2015wwa, Benbrik:2019zdp, Benbrik:2024hsf, Arhrib:2024nbj,
Benbrik:2023xlo,Benbrik:2025nfw,Benbrik:2022kpo,Benbrik:2024bxt,Arhrib:2024mbq,Arhrib:2024dou,Arhrib:2024tzm,Arhrib:2016rlj,Abouabid:2023mbu,Angelescu:2015uiz,Ghosh:2023xhs,Dermisek:2019vkc, Cingiloglu:2023ylm, Benbrik:2023quz, Benbrik:2026zjv}, identifying important implications for decay patterns and mass constraints.

Recently, the CMS Collaboration has conducted dedicated searches for singly produced vector-like $T$ quarks that decay exclusively into BSM final states such as $t\phi$, where $\phi$ denotes a neutral scalar boson and may correspond to $H$ or $A$ within the 2HDM+VLQ framework~\cite{CMS:2025zwi}. However, in the 2HDM+VLQ model, VLTs are not expected to decay exclusively into neutral scalar bosons; rather, the only BSM decay mode that can occur with a fully exclusive branching fraction is the charged Higgs channel for the $(T,B)$ doublet~\cite{Benbrik:2025kvz, Arhrib:2024tzm}. Therefore, the reported limits are not expected to constrain the mixing parameters of our model. Similar search strategies are anticipated to be extended to the pair-production regime in forthcoming analyses, which is expected to modify the current exclusion limits, particularly in scenarios where BSM decay modes dominate.

In this work, we investigate a VLB in both singlet and doublet representations. Our analysis reinterprets current pair-production exclusion limits using the inclusive branching-ratio rescaling method~\cite{Bhardwaj:2022nko, Benbrik:2025kvz}. We emphasize that the presence of non-standard decay channels such as $B \to Hb$, $B \to Ab$, and $B \to H^- t$ can substantially relax the existing mass bounds. Among the available experimental searches, we select those providing the most stringent upper limits in exclusive decay scenarios to ensure a consistent and reliable recasting procedure. This enables us to obtain updated exclusion reaches for scenarios in which these additional decay channels play a significant role. To evaluate how these non-standard modes influence the limits, we perform extensive parameter scans, exploring their dependence on $\tan\beta$, the scalar mass spectrum, and the relevant mixing angles. Furthermore, we assess how each individual BSM branching fraction contributes to the relaxation of the resulting upper limits.

The structure of the paper is as follows. Section~\ref{sec:Framework} introduces the theoretical framework. Section~\ref{sec:Constraints} summarizes the relevant theoretical and experimental constraints and outlines the methodology used to recast current LHC limits. Section~\ref{sec:Results} describes the setup of the numerical analysis and presents the main results. Finally, our conclusions are given in Section~\ref{sec:Conclusions}.

\section{Framework}
\label{sec:Framework}

In the 2HDM with a softly broken $Z_2$ symmetry, the scalar sector includes two complex $SU(2)_L$ doublets, $\Phi_1$ and $\Phi_2$, with the most general CP-conserving and gauge-invariant potential given by~\cite{Branco:2011iw,Gunion:1989we}:
\begin{eqnarray}
V\left( \Phi_1, \Phi_2 \right)  &=& m_{11}^2 \Phi_1^\dag \Phi_1 + m_{22}^2 \Phi_2^\dag \Phi_2 - m_{12}^2 \left( \Phi_1^\dag \Phi_2 + \Phi_2^\dag \Phi_1 \right) \nonumber \\
&+& \frac{\lambda_1}{2} \left( \Phi_1^\dag \Phi_1 \right)^2 + \frac{\lambda_2}{2} \left( \Phi_2^\dag \Phi_2 \right)^2
\nonumber \\
&+& \lambda_3 \left( \Phi_1^\dag \Phi_1 \right) \left( \Phi_2^\dag \Phi_2 \right) + \lambda_4 \left( \Phi_1^\dag \Phi_2 \right) \left( \Phi_2^\dag \Phi_1 \right)  \nonumber\\
&+& \frac{\lambda_5}{2} \left[ \left( \Phi_1^\dag \Phi_2 \right)^2 + \left( \Phi_2^\dag \Phi_1 \right)^2 \right],
\label{thdmV}
\end{eqnarray}
where all parameters are taken to be real.

Rotating to the so-called Higgs basis, only one linear combination of the two doublets acquires a vacuum expectation value (VEV),
\begin{eqnarray}
H_1 = \left( \begin{array}{c}
G^+ \\
\frac{v + \varphi^0_1 + i G^0}{\sqrt{2}} \\
\end{array} \right), \quad
H_2 = \left( \begin{array}{c}
H^+ \\
\frac{\varphi^0_2 + i A}{\sqrt{2}} \\
\end{array} \right),
\end{eqnarray}
where $v = \sqrt{v_1^2 + v_2^2} \simeq 246$ GeV is the electroweak scale, $G^0$ and $G^\pm$ are the Goldstone bosons, and $H^\pm$ is the charged Higgs. The CP-odd field $A$ and the CP-even fields $\varphi^0_{1,2}$ mix to give the physical neutral scalars $h$ and $H$:
\begin{eqnarray}
\left( \begin{array}{c}
h \\
H \\
\end{array} \right)
=
\left( \begin{array}{cc}
\sin(\beta - \alpha) & \cos(\beta - \alpha) \\
\cos(\beta - \alpha) & -\sin(\beta - \alpha) \\
\end{array} \right)
\left( \begin{array}{c}
\varphi_1^0 \\
\varphi_2^0 \\
\end{array} \right),
\end{eqnarray}
with $\tan\beta = v_2 / v_1$ and the mixing angle $\alpha$ diagonalizing the CP-even scalar mass matrix. In the alignment limit, $\sin(\beta - \alpha) \to 1$, the field $h$ behaves like the SM Higgs boson.             

VLQs are heavy fermions whose left- and right-handed components transform identically under the electroweak gauge group. They appear in various BSM scenarios, such as extra-dimensional models~\cite{Gherghetta:2000qt}, composite Higgs theories~\cite{Contino:2006qr, Agashe:2004rs}, and GUTs~\cite{Hewett:1988xc}, and allow for gauge-invariant mass terms without requiring electroweak symmetry breaking. Their representations under $SU(3)_C \times SU(2)_L \times U(1)_Y$ include:
\begin{align}
& T^0_{L,R}, \quad B^0_{L,R} && \text{(singlets)} \,, \notag \\
& (X, T^0)_{L,R}, \quad (T^0, B^0)_{L,R}, \quad (Y, B^0)_{L,R} && \text{(doublets)} \,, \notag \\
& (X, T^0, B^0)_{L,R}, \quad (T^0, B^0, Y)_{L,R} && \text{(triplets)} \,.
\end{align}
Here, the superscript $0$ indicates weak eigenstates, which will be omitted when the context is clear. The $B$-type quark carries electric charge $Q = -1/3$ and mixes with the SM bottom quark after electroweak symmetry breaking.

In this work, we focus on the VLB as either a singlet or a member of a $(T, B)$ or $(B, Y)$ doublet. The presence of $B^0_{L,R}$ leads to a modification of the down-type quark sector, yielding four mass eigenstates: $d$, $s$, $b$, and $B$. The mixing is primarily with the third generation, due to stringent constraints from LEP measurements of $R_b$~\cite{Aguilar-Saavedra:2002phh}. The mixing between $b^0$ and $B^0$ is parametrized by:
\begin{eqnarray}
\left( \begin{array}{c}
b_{L,R} \\
B_{L,R} \\
\end{array} \right)
=
\left( \begin{array}{cc}
\cos\theta_{L,R}^d & -\sin\theta_{L,R}^d e^{i \phi_d} \\
\sin\theta_{L,R}^d e^{-i \phi_d} & \cos\theta_{L,R}^d \\
\end{array} \right)
\left( \begin{array}{c}
b^0_{L,R} \\
B^0_{L,R} \\
\end{array} \right),
\label{ec:mixd}
\end{eqnarray}
where $\theta_{L,R}^d$ are the left- and right-handed mixing angles, and $\phi_d$ is a CP-violating phase, which we neglect in this work.

The Yukawa sector in the Higgs basis includes:
\begin{equation}
-\mathcal{L}_Y \supset y^u \bar{Q}^0_L \tilde{H}_2 u^0_R + y^d \bar{Q}^0_L H_1 d^0_R + M_u^0 \bar{u}_L^0 u_R^0 + M_d^0 \bar{d}_L^0 d_R^0 + \text{h.c.},
\end{equation}
with $u_R^0 = (u_R, c_R, t_R, T_R)$ and $d_R^0 = (d_R, s_R, b_R, B_R)$. The VLB mass matrix takes the form:
\begin{eqnarray}
\mathcal{L}_\text{mass} = - \left( \begin{array}{cc}
\bar{b}_L^0 & \bar{B}_L^0 \\
\end{array} \right)
\left( \begin{array}{cc}
y_{33}^d \frac{v}{\sqrt{2}} & y_{34}^d \frac{v}{\sqrt{2}} \\
y_{43}^d \frac{v}{\sqrt{2}} & M^0 \\
\end{array} \right)
\left( \begin{array}{c}
b_R^0 \\
B_R^0 \\
\end{array} \right) + \text{h.c.},
\label{ec:Lmass}
\end{eqnarray}
where $M^0$ is a bare vector-like mass term, and $y_{ij}^d$ are Yukawa couplings. Diagonalization proceeds via a bi-unitary transformation:
\begin{equation}
U_L^d \mathcal{M}^d (U_R^d)^\dagger = \mathcal{M}^d_\text{diag} \,.
\label{ec:diag}
\end{equation}

The mixing angles obey the relations:
\begin{eqnarray}
\tan 2 \theta_L^d &=& \frac{\sqrt{2} |y_{34}^d| v M^0}{(M^0)^2 - \frac{1}{2} v^2(|y_{33}^d|^2 + |y_{34}^d|^2)} \quad \text{(singlets, triplets)}, \notag \\
\tan 2 \theta_R^d &=& \frac{\sqrt{2} |y_{43}^d| v M^0}{(M^0)^2 - \frac{1}{2} v^2(|y_{33}^d|^2 + |y_{43}^d|^2)} \quad \text{(doublets)}.
\label{ec:angle1}
\end{eqnarray}
Additionally,
\begin{eqnarray}
\tan \theta_R^q &=& \frac{m_q}{m_Q} \tan \theta_L^q \quad \text{(singlets, triplets)}, \notag \\
\tan \theta_L^q &=& \frac{m_q}{m_Q} \tan \theta_R^q \quad \text{(doublets)}.
\label{ec:rel-angle1}
\end{eqnarray}

In the alignment limit of the 2HDM, the interactions between the VLB and the additional scalar states are described by:
\begin{align}
\mathcal{L}_{H} &= -\frac{g m_B}{2 M_W} \overline{b} \left( Y^L_{HbB} P_L + Y^R_{HbB} P_R \right) B H + \text{h.c.}, \label{eq:L_H} \\
\mathcal{L}_{A} &= i \frac{g m_B}{2 M_W} \overline{b} \left( Y^L_{AbB} P_L - Y^R_{AbB} P_R \right) B A + \text{h.c.}, \label{eq:L_A} \\
\mathcal{L}_{H^-} &= -\frac{g m_B}{\sqrt{2} M_W} \overline{B} \left( \cot\beta Z^L_{Bt} P_L + \tan\beta Z^R_{Bt} P_R \right) b H^- + \text{h.c.}, \label{eq:L_Hplus}
\end{align}
where $Y^{L,R}$ and $Z^{L,R}$ encode the chiral couplings of the VLB to the neutral and charged Higgs bosons. Their explicit expressions and the corresponding partial widths are provided in~\ref{sec:cpl}. Interactions involving purely heavy or purely light fermions are discussed in detail in Ref.~\cite{Arhrib:2024tzm}.

\section{Theoretical and Experimental Constraints}
\label{sec:Constraints}

We impose a set of theoretical and experimental requirements on the model parameter space to ensure consistency with perturbative unitarity, vacuum stability, electroweak precision data, and collider bounds.

\subsection*{Theoretical constraints}
Tree-level theoretical constraints from perturbative unitarity, perturbativity, and vacuum stability are imposed on the scalar potential of the 2HDM sector. Since VLQs do not directly contribute to the scalar potential, these conditions remain unaltered at tree level. Their effects enter only at loop level through corrections to electroweak precision observables and via their Yukawa interactions~\cite{Cingiloglu:2023ylm}.

\begin{itemize}
\item \textbf{Unitarity:} The $S$-wave amplitudes for scalar--scalar, scalar--gauge, and gauge--gauge scattering must satisfy perturbative unitarity at high energies~\cite{Kanemura:1993hm}.

\item \textbf{Perturbativity (scalar sector):} All quartic couplings in the scalar potential are required to obey $|\lambda_i| < 8\pi$ for $i=1,\dots,5$~\cite{Branco:2011iw}, ensuring the validity of the perturbative expansion.

\item \textbf{Vacuum stability:} The potential must be bounded from below in any field direction. This leads to the conditions~\cite{Deshpande:1977rw, Barroso:2013awa}:
\begin{align}
&\lambda_1 > 0,\quad \lambda_2 > 0,\quad \lambda_3 > -\sqrt{\lambda_1 \lambda_2}, \notag \\
&\lambda_3 + \lambda_4 - |\lambda_5| > -\sqrt{\lambda_1 \lambda_2}.
\end{align}

\item \textbf{Electroweak precision observables (EWPOs):} The oblique parameters $S$ and $T$~\cite{Grimus:2007if} are constrained at the 95\% confidence level (CL) according to the global electroweak fit, assuming $U = 0$~\cite{ParticleDataGroup:2020ssz}:
\begin{align}
S = 0.05 \pm 0.08,\quad T = 0.09 \pm 0.07,\quad \rho_{ST} = 0.92.
\end{align}
In the presence of VLQs, the total contributions are evaluated as $\chi^2(S_{\text{2HDM}} + S_{\text{VLQ}},\,T_{\text{2HDM}} + T_{\text{VLQ}})$ and tested against the above bounds. The VLQ-induced corrections to EWPOs follow the analytic results of Ref.~\cite{Arhrib:2024tzm}. Requiring consistency with the 95\% CL allowed region significantly constrains the VLQ mixing parameters, leading to $s_R^u,\, s_R^d \lesssim 0.2$ throughout the viable parameter space. All constraints are implemented using a modified version of \texttt{2HDMC-1.8.0}~\cite{Eriksson:2009ws}, incorporating VLQ contributions as discussed in Refs.~\cite{Benbrik:2022kpo, Abouabid:2023mbu}.

\end{itemize}

\subsection*{Experimental constraints}

\begin{itemize}
\item \textbf{Searches for additional Higgs bosons:} 
Direct searches for heavy neutral ($H$, $A$) and charged ($H^\pm$) Higgs bosons impose significant constraints on the 2HDM parameter space. For neutral scalars, LHC analyses probe production via gluon-gluon fusion and $b$-associated production, with decay channels including $\tau^+\tau^-$~\cite{ATLAS:2020zms, CMS:2022goy}, $ZA/H \to \ell\ell bb$ or $\ell\ell WW$~\cite{ATLAS:2020gxx}, $\gamma\gamma$~\cite{ATLAS:2021uiz}, and $t\bar{t}$~\cite{CMS:2019pzc}. Charged Higgs searches primarily target $H^+ \to tb$~\cite{ATLAS:2021upq, CMS:2020imj} and $H^+ \to \tau^+\nu_\tau$~\cite{ATLAS:2018gfm, CMS:2019bfg}. Within the 2HDM+VLQ framework, VLQs can modify Higgs production and decay rates through light-light couplings~\cite{Arhrib:2024tzm, Arhrib:2024dou}\footnote{A detailed analysis of the impact of these couplings is beyond the scope of the present work.}. These constraints are implemented using \texttt{HiggsBounds-6} within the \texttt{HiggsTools} framework~\cite{Bechtle:2008jh, Bechtle:2011sb, Bechtle:2013wla, Bechtle:2015pma, Bahl:2022igd}, which systematically tests each parameter point against exclusion limits from LEP, Tevatron, and the LHC.

\item \textbf{SM-like Higgs measurements:}
Compatibility with the observed 125~GeV Higgs boson is evaluated using \texttt{HiggsSignals-3} within the \texttt{HiggsTools} framework~\cite{Bechtle:2020uwn, Bechtle:2020pkv}, requiring $\Delta \chi^2 \leq 6.18$ at 95\% CL across 159 signal-strength measurements. In the 2HDM+VLQ setup, VLQs contribute to loop-induced processes such as $h \to gg$ and $h \to \gamma\gamma$. Previous studies have shown that these effects typically reduce $\mathcal{BR}(h \to gg)$ and $\mathcal{BR}(h \to \gamma\gamma)$ by up to approximately 10\% and 3\%, respectively~\cite{Arhrib:2024tzm}, which remains well within current experimental uncertainties~\cite{ATLAS:2021vrm}.

\item \textbf{$b \to s\gamma$ constraint:} In the 2HDM-II, the radiative transition $b \to s\gamma$ sets a strong lower bound $m_{H^\pm} \gtrsim 580$~GeV. In the presence of VLQs, this limit may be significantly relaxed via loop-induced cancellations. For instance, in the ($T, B$) doublet case, viable configurations exist with $m_{H^\pm} \sim 360$~GeV depending on the mixing~\cite{Benbrik:2022kpo}. In our analysis, we conservatively take $m_{H^\pm} \geq 600$~GeV.

\item \textbf{LHC Constraints on VLQs:}
Constraints on the VLB from LHC searches are implemented by requiring $\sigma_{\text{theo}}/\sigma_{\text{obs}} < 1$, following the procedure of Ref.~\cite{Benbrik:2024fku}. Single-production searches primarily constrain couplings to SM final states through channels such as $bqq\ell\nu$ and $b\ell\nu qq$~\cite{CMS:2018dcw, CMS:2021mku}.\footnote{These limits assume $\mathcal{BR}(B \to \text{BSM}) \approx 0$ and therefore apply only when decays into heavy Higgs bosons are negligible.} Current pair-production bounds are derived under the same assumption of exclusive decays into SM channels ($B \to Wt$, $Zb$, $hb$). In this work, we reinterpret these bounds by incorporating additional decay modes into heavy Higgs states ($H$, $A$, $H^\pm$), which reduce the branching fractions into SM final states and consequently weaken the extracted mass limits.

\end{itemize}

\section{Recasting LHC Bounds}
\label{sec:lhc}
At the LHC, VLQs can be produced through two main mechanisms: pair production and single production. Pair production, driven by QCD interactions, is largely model-independent, as its cross section depends primarily on the VLQ mass. In contrast, single production proceeds via EW interactions, making it more sensitive to the couplings between VLQs and SM quarks.

Current LHC searches set stringent limits on pair-produced VLBs under the assumption of exclusive decays into SM final states. As reported in~\cite{Benbrik:2024fku}, these analyses exclude masses up to about $m_B \sim 1.5$~TeV. To account for possible non-standard decay channels, we consider the most constraining ATLAS and CMS results~\cite{CMS:2024bni, ATLAS:2022tla}. The corresponding production cross sections, used to derive the mass limits, are computed at NNLO+NNLL accuracy in QCD with \texttt{Top++} employing the MSTW2008nnlo PDF set~\cite{Martin:2009iq,Martin:2009bu,Martin:2010db}. The recasting is performed following the model-independent strategy proposed in~\cite{Bhardwaj:2022nko}, which enables reinterpretation for arbitrary BR configurations.\\

For the singlet scenario, the BRs approximately satisfy:
\begin{equation}
\mathcal{BR}(B \to Zb) \simeq \mathcal{BR}(B \to hb) \simeq \tfrac{1}{2}\,\mathcal{BR}(B \to Wt),
\end{equation}
while in the doublet case:
\begin{equation}
\mathcal{BR}(B \to Zb) \simeq \mathcal{BR}(B \to hb), \quad \mathcal{BR}(B \to Wt) \simeq 0,
\label{equ:dblt}
\end{equation}
valid at the TeV scale for small mixing. The total BR into SM final states satisfies:
\begin{equation}
\mathcal{BR}_{\text{SM}} = 1 - \mathcal{BR}_{\text{BSM}},
\label{eq:brsum}
\end{equation}
with $\mathcal{BR}_{\text{BSM}} \equiv \mathcal{BR}(B \to Hb) + \mathcal{BR}(B \to Ab) + \mathcal{BR}(B \to H^-t)$.\\

The inclusive BR for an SM channel $i$ in $B\bar{B}$ production is given by:
\begin{equation}
\mathcal{B}^{\text{inc}}_i = \mathcal{BR}_i^2 + 2 \sum_{j \neq i} \mathcal{BR}_i\,\mathcal{BR}_j = \mathcal{BR}_i (2 - \mathcal{BR}_i),
\label{eq:br_inc}
\end{equation}
capturing both symmetric and mixed final states.

\begin{figure}[htp]
\centering
\includegraphics[width=7cm]{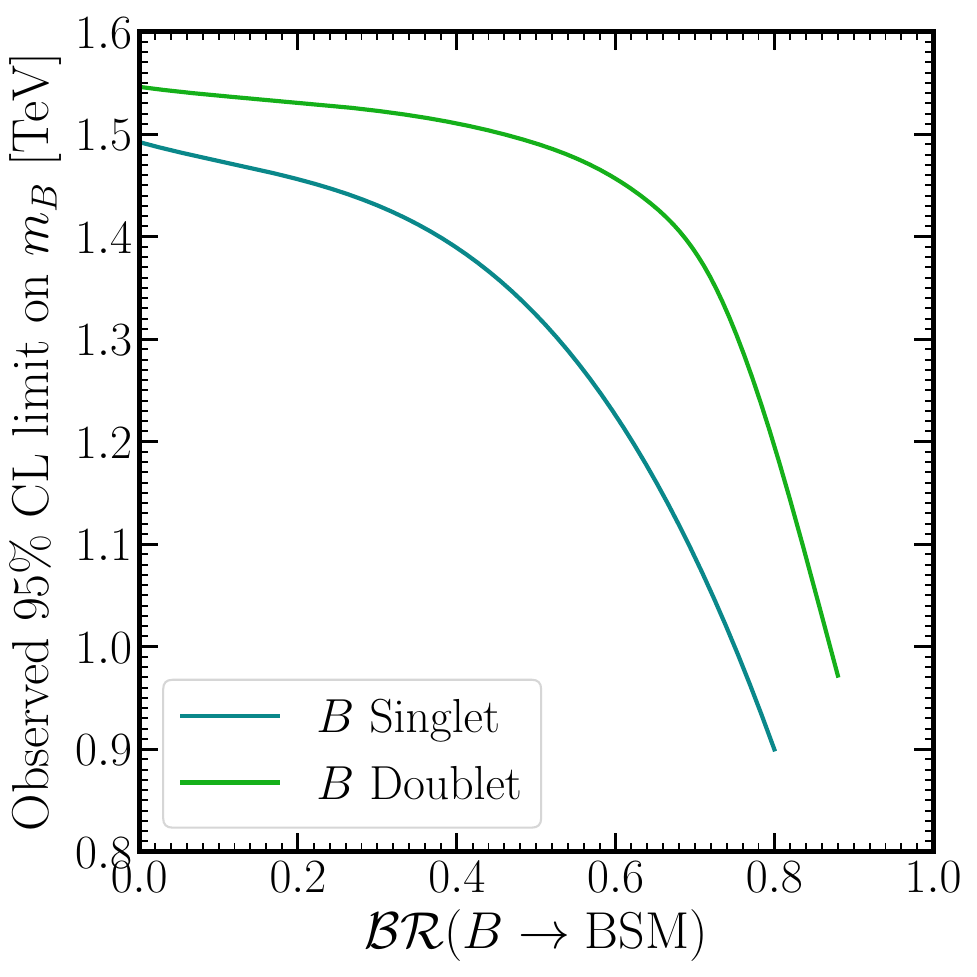}
\caption{Recast LHC exclusions on $m_B$ as a function of $\mathcal{BR}_{\text{BSM}}$. Blue (green) line: singlet (doublet).}
\label{fig:bsm}
\end{figure}

By scanning over $\mathcal{BR}_{\text{BSM}} \in [0,1]$, we rescale the effective signal cross section and extract the corresponding exclusion limits assuming unchanged selection efficiencies and neglecting potential overlaps between exotic final states and existing signal regions, as shown in Fig.~\ref{fig:bsm}. 
Recasting the exclusive limits in the $\mathcal{BR}_{\text{BSM}} = 0$ limit reproduces excluded masses of approximately $m_B \sim 1.5$~TeV for the singlet and $m_B \sim 1.55$~TeV for the doublet. As $\mathcal{BR}_{\text{BSM}}$ increases, the suppression of SM decay channels progressively weakens the exclusion reach. For $\mathcal{BR}_{\text{BSM}} \sim 0.80$ (singlet) and $\sim 0.89$ (doublet), the lower bounds decrease to $m_B \sim 0.94$~TeV and $m_B \sim 0.98$~TeV, respectively. Beyond these values, the SM branching fractions become too small to sustain meaningful constraints, and conventional searches lose sensitivity when $B \to \text{BSM}$ decays dominate. As $\mathcal{BR}_{\text{BSM}}$ increases, the SM channels are suppressed, weakening the exclusions. For $\mathcal{BR}_{\text{BSM}} \sim 0.80$ (singlet) and $\sim 0.89$ (doublet), the limits drop to $m_B \sim 0.94$~TeV and $m_B \sim 0.98$~TeV, respectively.
The disappearance of the exclusion contours for $\mathcal{BR}(B \to \mathrm{BSM}) > 0.8$ (0.9) signals that the SM decay modes become too suppressed to provide meaningful constraints. In this regime, SM-based limits no longer apply, effectively allowing the entire VLB mass range since no mass exclusion can be derived when $B \to \mathrm{BSM}$ decays dominate.

\section{Results and Discussion}
\label{sec:Results}
We investigate the phenomenological implications of the 2HDM-II extended by a VLB quark, considering both singlet and doublet representations. In particular, we examine how the presence of BSM decay modes, namely $B \to Hb$, $B \to Ab$, and $B \to H^-t$, modifies the sensitivity of LHC searches, which are typically optimized for SM final states. Our scan covers the parameter space:
\begin{align*}
&m_B \in [0.8, 2]~\text{TeV}, \quad \theta_{L/R}^{u/d} \in [-\frac{\pi}{6}, \frac{\pi}{6}], \quad \tan\beta \in [0.5, 10],\\
&m_{H,A} \in [130, 800]~\text{GeV}, \quad m_{H^\pm} \in [600, 1000]~\text{GeV}
\end{align*}

The ranges $\theta \in [-\frac{\pi}{6}, \frac{\pi}{6}]$ and $\tan\beta \in [0.5, 10]$ are adopted to ensure that the majority of the parameter space is not excluded by electroweak precision tests (STU) or existing collider constraints.

Once kinematically open, the BSM decay modes dominate the partial widths of the VLB, significantly reducing the BRs into $Wt$, $Zb$, and $hb$. This suppression of conventional final states leads to a decreased efficiency in current searches, and consequently, to weaker exclusion bounds on $m_B$. We quantify this behavior through the inclusive branching ratio into non-SM final states:
\begin{equation}
\mathcal{BR}_{\text{BSM}} \equiv \mathcal{BR}(B \to Hb) + \mathcal{BR}(B \to Ab) + \mathcal{BR}(B \to H^-t),
\end{equation}
and study its correlation with the excluded mass bounds in the $(m_B,\, \mathcal{BR}_{\text{BSM}})$ plane.

We find that the impact of $\mathcal{BR}_{\text{BSM}}$ is particularly pronounced in the doublet case, where the SM-like decay pattern dominates for $\mathcal{BR}_{\text{BSM}} \to 0$. As $\mathcal{BR}_{\text{BSM}}$ increases, the exclusion limits drop considerably for both representations, emphasizing the necessity of including these non-standard final states in dedicated collider analyses.

\subsection{2HDM-II with VLB Singlet}

\begin{figure*}[h!]
\centering
\includegraphics[width=15cm]{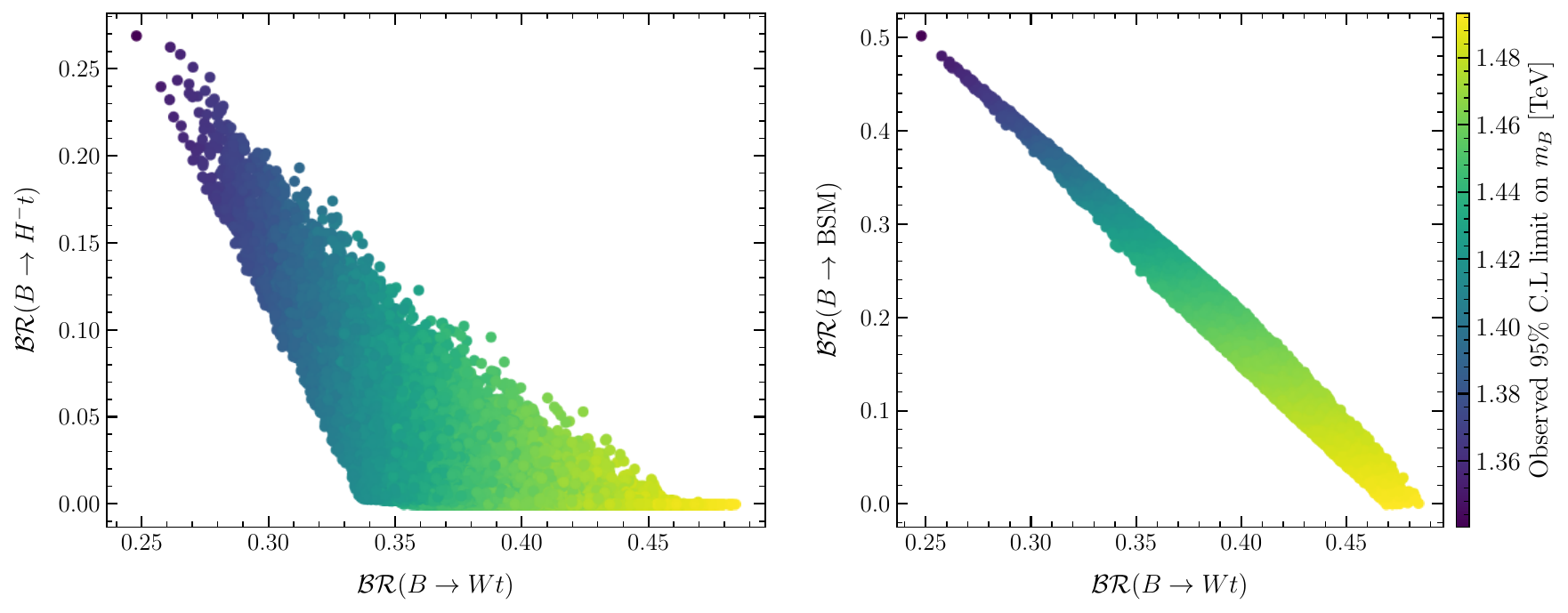}
\caption{Left: $\mathcal{BR}(B \to Wt)$ versus $\mathcal{BR}(B \to H^-t)$; Right: total $\mathcal{BR}_{\text{BSM}} \equiv \mathcal{BR}(B \to Ab) + \mathcal{BR}(B \to Hb) + \mathcal{BR}(B \to H^-t)$ versus $\mathcal{BR}(B \to Wt)$. The color bar shows the recast lower limit on $m_B$ in TeV for the 2HDM-II + VLB singlet scenario.}
\label{fig:lim_sing}
\end{figure*}

In Fig.~\ref{fig:lim_sing}, we present the recast exclusion limits on $m_B$ in the 2HDM-II + VLB singlet scenario. The left panel shows the interplay between $\mathcal{BR}(B \to Wt)$ and $\mathcal{BR}(B \to H^-t)$, while the right panel displays $\mathcal{BR}_{\text{BSM}}$ versus $\mathcal{BR}(B \to Wt)$. The color bar indicates the excluded $m_B$ values. We observe that $\mathcal{BR}(B \to H^-t)$ can reach up to $\sim 27\%$ when $\mathcal{BR}(B \to Wt)$ is reduced to $\sim 27\%$, with $\mathcal{BR}_{\text{BSM}}$ attaining values as high as $50\%$. This translates into a relaxation of the $m_B$ bound from $\sim 1.5$~TeV to $\sim 1.34$~TeV.

\begin{figure*}[h!]
\centering
\includegraphics[width=15cm, height=210pt]{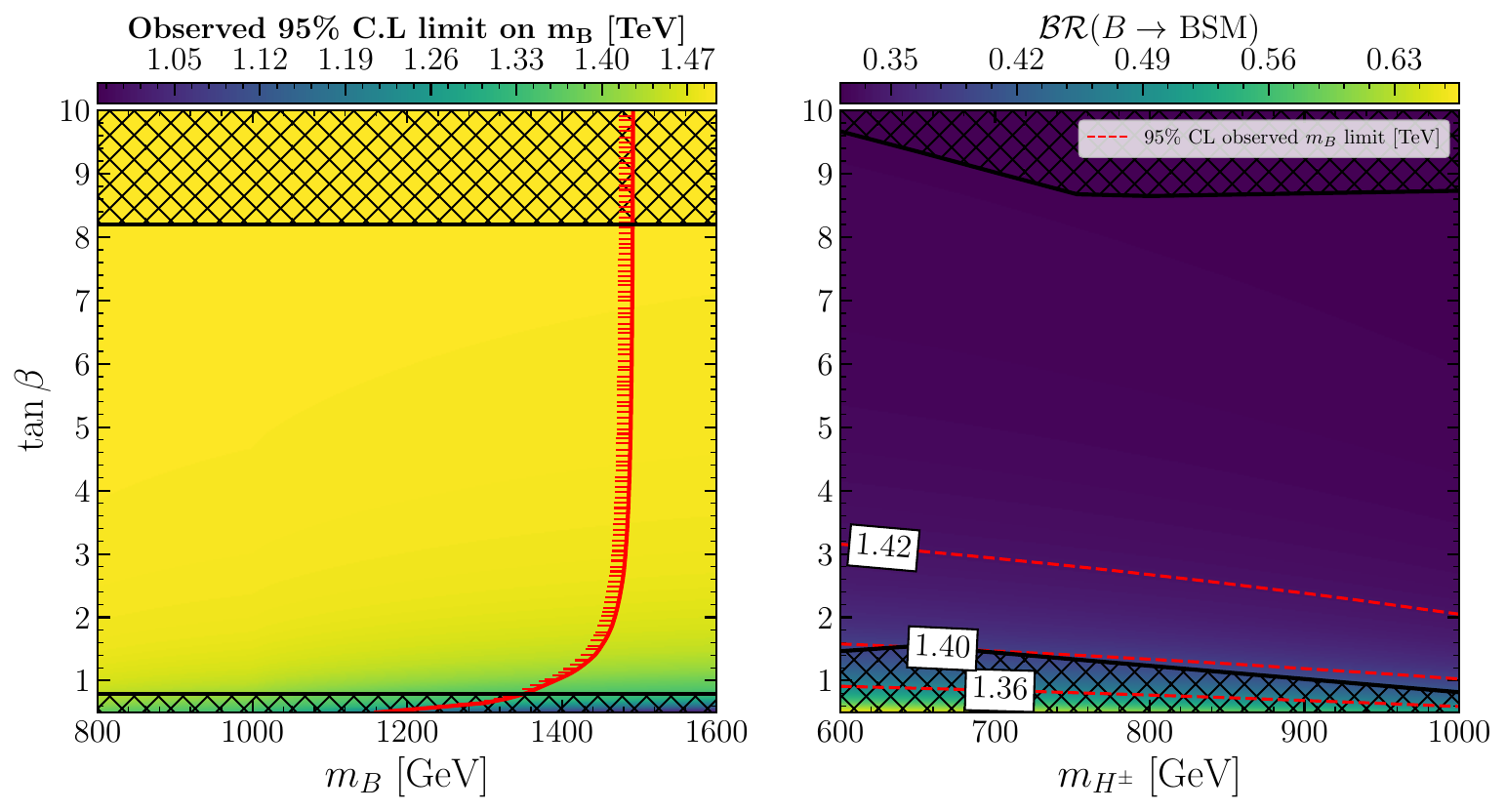}
\caption{The left panel shows the exclusion contours in the $(m_B, \tan\beta)$ plane, while the right panel presents the $\mathcal{BR}(B \to \text{BSM})$ projected in the $(m_{H^\pm}, \tan\beta)$ plane within the 2HDM-II + VLB singlet scenario. In the right plot, the red dashed curves indicate the 95\% CL exclusion. The lower shaded region is excluded by the ATLAS search for $H^+ \to tb$\cite{ATLAS:2021upq}, whereas the upper shaded region is excluded by the ATLAS $\tau\tau$ search\cite{ATLAS:2020zms}. The parameter scan is defined as $m_A = m_H = 500~\text{GeV}$, $m_{H^\pm} = 832~\text{GeV}$, and $s_L = 0.1$ for the left panel. The right panel uses the same parameters, except for $m_B = 1.5~\text{TeV}$.}
\label{fig:exclu_singl}
\end{figure*}

To illustrate the exclusion sensitivity in physical parameter planes, we select a benchmark configuration and project the constraints onto the $(m_B, \tan\beta)$ plane in the left panel and the branching ratio $\mathcal{BR}(B \to \text{BSM})$ onto the $(m_{H^\pm}, \tan\beta)$ plane in the right panel, as shown in Fig.~\ref{fig:exclu_singl}. In the left panel, we scan $m_B \in [0.8, 1.6]$~TeV and $\tan\beta \in [0.5, 10]$. In the right panel, we scan $m_{H^\pm} \in [600, 1000]$~GeV and $\tan\beta \in [0.5, 10]$, with fixed parameters $m_A = m_H = 500$~GeV, $s_L = 0.1$, and $s_{\beta - \alpha} = 1$. We set $m_{H^\pm} = 832$~GeV for the left panel and $m_B = 1.5$~TeV for the right panel. The lower shaded region is excluded by the ATLAS $H^+ \to tb$ search~\cite{ATLAS:2021upq}, while the upper shaded region is excluded by the ATLAS $\tau\tau$ search~\cite{ATLAS:2020zms}. The dashed red contour denotes the 95\% CL limit from the recast analysis. Near $\tan\beta \sim 1$, the exclusion extends to $m_B \sim 1.4$~TeV. For $\tan\beta \lesssim 1$, the exclusion weakens slightly due to the $1/\tan^2\beta$ scaling of $\mathcal{BR}(B \to H^- t)$. At larger $\tan\beta$, SM branching ratios dominate, stabilizing $m_B$ at approximately 1.48~TeV. In the right panel, contour lines illustrate the observed limit as a function of $m_{H^\pm}$ and $\tan\beta$. The exclusion similarly weakens for $\tan\beta \lesssim 1$ due to enhanced BSM branching ratios, with minimal dependence on $m_{H^\pm}$.

\begin{figure*}[h!]
\centering
\includegraphics[width=15cm]{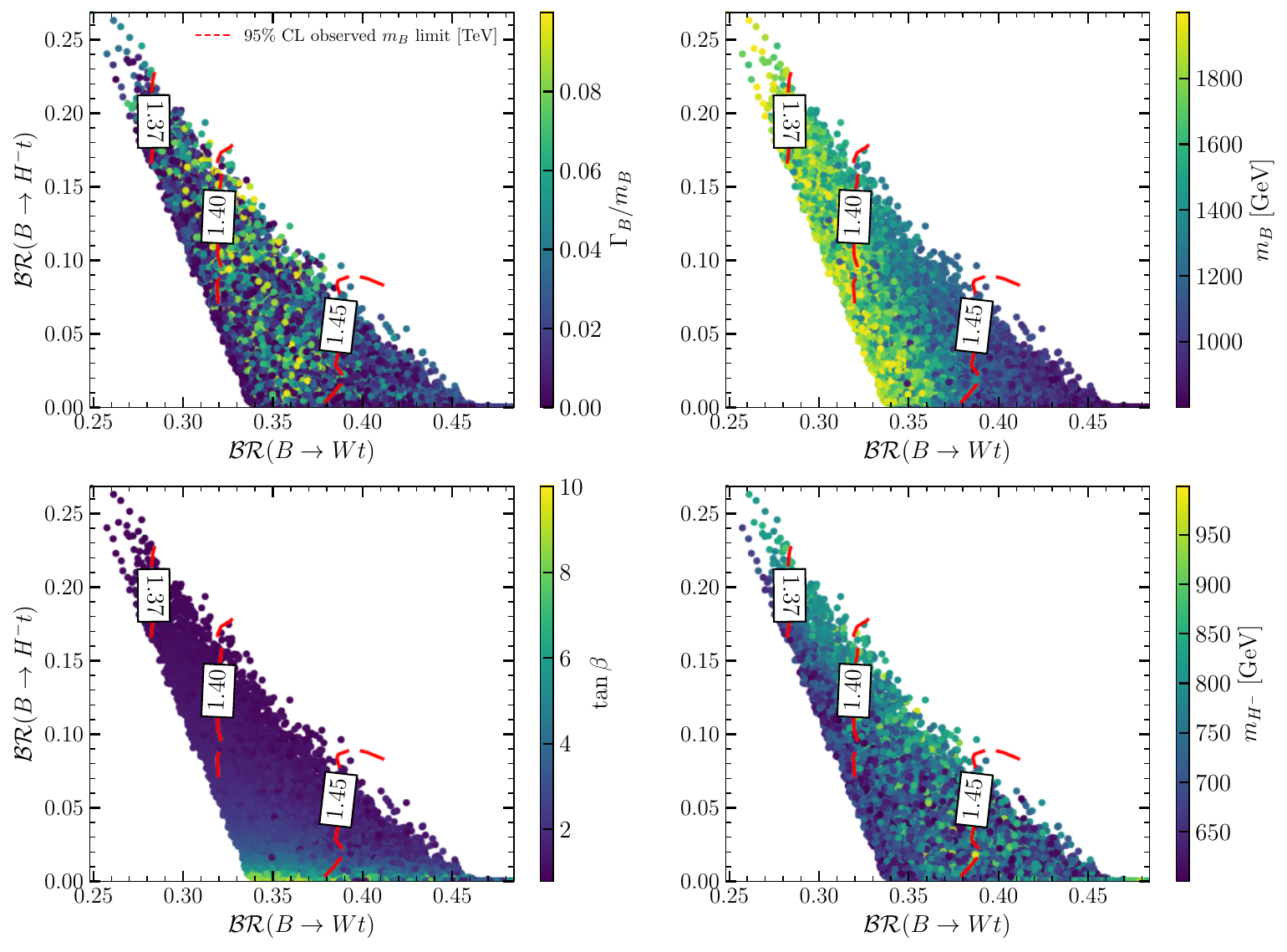}
\caption{Distributions of $\mathcal{BR}(B \to Wt)$ vs. $\mathcal{BR}(B \to H^-t)$ for the VLB singlet scenario. Color-coded by: $\Gamma_B/m_B$ (top left), $m_B$ (top right), $\tan\beta$ (bottom left), and $m_{H^\pm}$ (bottom right). Red dashed contours denote the observed $m_B$ limits.}
\label{fig:sing_params}
\end{figure*}
\begin{figure*}[h!]
\centering
\includegraphics[width=15.5cm]{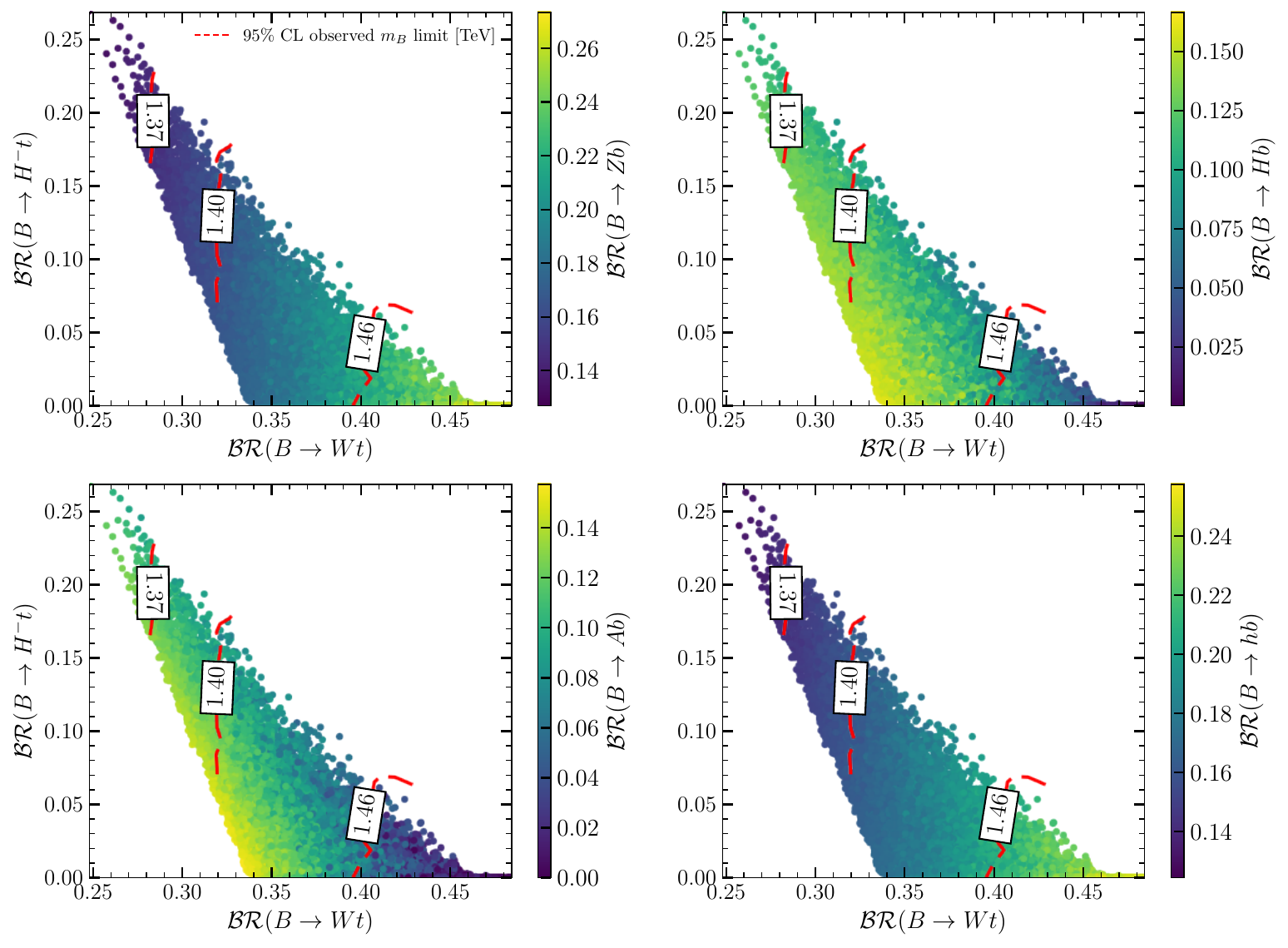}
\caption{$\mathcal{BR}(B \to H^-t)$ vs. $\mathcal{BR}(B \to Wt)$ colored by: $\mathcal{BR}(B \to Zb)$ (top left), $\mathcal{BR}(B \to Hb)$ (top right), $\mathcal{BR}(B \to Ab)$ (bottom left), and $\mathcal{BR}(B \to hb)$ (bottom right) in the 2HDM-II + VLB singlet scenario.}
\label{fig:g_params}
\end{figure*}
Fig.~\ref{fig:sing_params} presents the dependence of the decay modes ${\cal BR}(B \to Wt)$ and ${\cal BR}(B \to H^- t)$ on key model parameters: the relative width $\Gamma_B/m_B$ (upper left), the VLB mass $m_B$ (upper right), $\tan\beta$ (lower left), and the charged Higgs mass $m_{H^\pm}$ (lower right). The red contours correspond to the recast $m_B$ exclusion limits. The ratio $\Gamma_B/m_B$ increases with the enhancement of ${\cal BR}(B \to H^- t)$. The $\mathcal{BR}(B \to H^- t)$ increases with $m_B$ due to $m_B^3$ scaling. The bottom-left panel confirms enhanced $\mathcal{BR}(B \to H^- t)$ at low $\tan\beta$, driven by $1/\tan^2\beta$ dependence. The bottom-right panel shows minimal dependence of $\mathcal{BR}(B \to H^- t)$ on $m_{H^\pm}$, consistent with Fig.~\ref{fig:exclu_singl}.

The correlation between $\mathcal{BR}(B \to Wt)$ and $\mathcal{BR}(B \to H^- t)$ is presented in Fig.~\ref{fig:g_params}. The color scale indicates the branching ratios $\mathcal{BR}(B \to Zb)$ (upper left), $\mathcal{BR}(B \to Hb)$ (upper right), $\mathcal{BR}(B \to Ab)$ (lower left), and $\mathcal{BR}(B \to hb)$ (lower right). In the absence of BSM decays, the SM-like branching ratio pattern for $(Zb, hb, Wt)$ approaches $\approx (0.25, 0.25, 0.5)$. The $Wt$ channel dominates, achieving a branching fraction of up to 50\%, while the SM decays $Zb$ and $hb$ reach approximately 28\% and 26\%, respectively. Among BSM channels, $H^- t$, $Hb$, and $Ab$ attain branching fractions of up to approximately 24\%, 16\%, and 16\%, respectively. The observed $m_B$ limit, indicated by the red lines, decreases from 1.46~TeV to 1.37~TeV, particularly in regions where the BSM branching ratios satisfy $\mathcal{BR}(B \to H^- t) > 20\%$ and $\mathcal{BR}(B \to (H/A)b) > 12\%$.

Finally, Table~\ref{tab:BpT} provides a set of benchmark points (BPs) chosen to illustrate distinct decay topologies. They are selected for yielding the largest BSM branching ratios and for satisfying the condition $m_B$ above the observed mass bound, ensuring their viability within the model.
For each BP, we report the relevant input parameters, branching ratios, total width, and the corresponding observed limit.

\begin{table*}[htpb!]
\centering
\renewcommand{\arraystretch}{0.8}
\setlength{\tabcolsep}{30pt}
\begin{adjustbox}{max width=\textwidth}
	\begin{tabular}{lccccc}
		\toprule\toprule
		Parameter & BP$_1$ & BP$_2$ & BP$_3$ & BP$_4$ & BP$_5$ \\
		\midrule
		\multicolumn{6}{c}{2HDM-II + VLB Inputs (masses in GeV)} \\
		\midrule
		$m_h$ & 125.1 & 125.1 & 125.1 & 125.1 & 125.1 \\
		$m_H$ & 354.105 & 543.995 & 486.302 & 427.817 & 445.770 \\
		$m_A$ & 497.977 & 600.544 & 728.140 & 503.416 & 587.960\\
		$m_{H^\pm}$ & 625.483 & 774.267 & 661.496 & 670.487 & 704.013 \\
		$\tan\beta$ & 1.202 & 1.031 & 1.273 & 1.284 & 1.240 \\
		$m_B$ & 1385.046 & 1398.283 & 1422.399 & 1396.700 & 1440.090 \\
		$\sin\theta_L$ & 0.106 & 0.226 & $-$0.011 & 0.105 & 0.179 \\
		\midrule
		\multicolumn{6}{c}{Branching Ratios (\%)} \\
		\midrule
		$\text{BR}(B \to Wt)$ & 29.992 & 31.821 & 31.973 & 31.036 & 31.630 \\
		$\text{BR}(B \to Zb)$ & 15.711 & 15.980 & 16.896 & 16.248 & 16.161 \\
		$\text{BR}(B \to hb)$ & 15.292 & 15.559 & 16.459 & 15.819 & 15.750 \\
		$\text{BR}(B \to Hb)$ & 13.579 & 11.386 & 13.037 & 13.201 & 13.072 \\
		$\text{BR}(B \to Ab)$ & 11.784 & 10.515 & 9.102 & 12.169 & 11.102 \\
		$\text{BR}(B \to H^-t)$ & 13.639 & 14.737 & 12.530 & 11.524 & 12.281 \\
		\midrule
        \multicolumn{6}{c}{Total width $\Gamma_B$ [GeV]} \\\toprule
		$\Gamma_B$ (GeV) & 31.135 & 137.632 & 0.370 & 30.301 & 95.113 \\
        \toprule
        \multicolumn{6}{c}{Observed $m_B$ limit [GeV]} \\\toprule
            $m_B^{\text{obs}}$ & 1383.676 & 1394.081 & 1402.752 & 1392.956 & 1394.884 \\
		\bottomrule\bottomrule
	\end{tabular}
\end{adjustbox}
\caption{Representative benchmark points for the 2HDM-II + VLB singlet scenario, showing scalar masses, mixing, BRs, total width, and observed mass limit at 95\% CL ($m^{\text{obs}}_B$).}
\label{tab:BpT}
\end{table*}

\subsection{2HDM-II with Doublet ($T, B$)}
\begin{figure*}[h!]
\centering
\includegraphics[width=15cm]{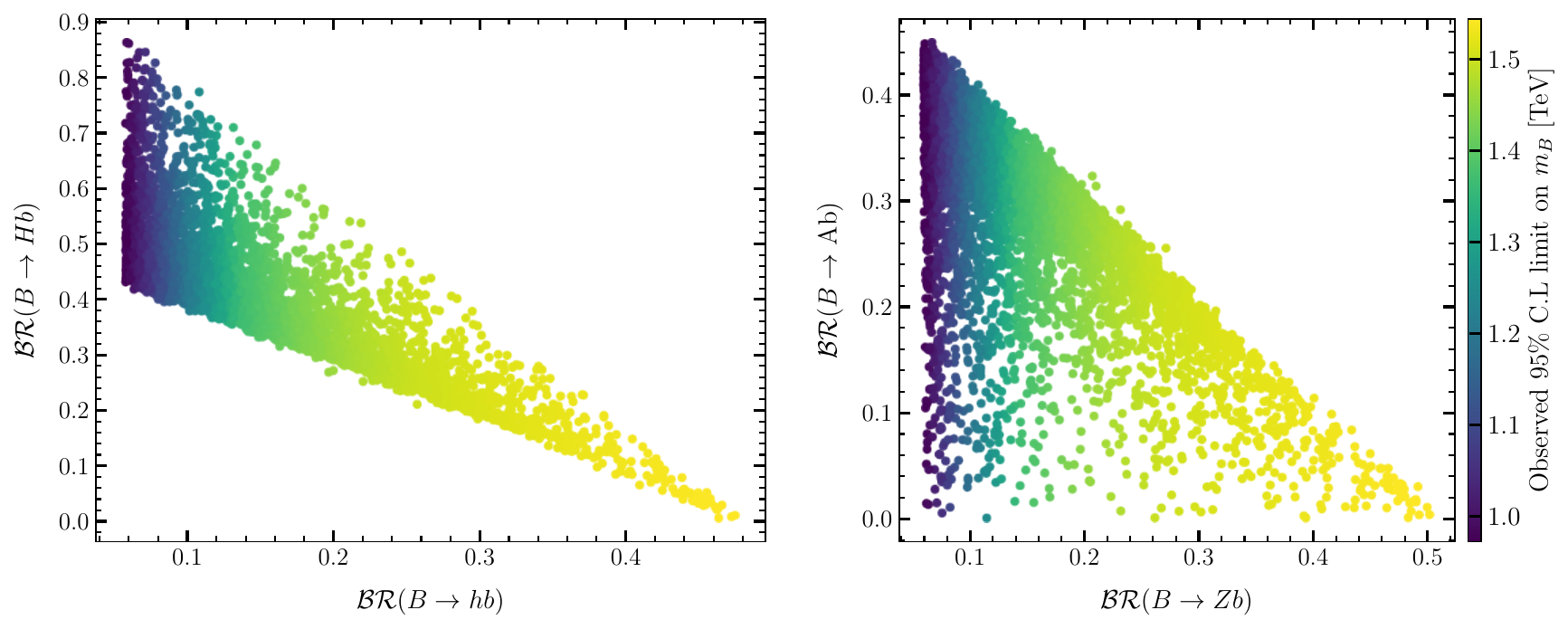}
\caption{Observed 95\% CL lower limit on $m_B$ in the $(\mathcal{BR}(B \to h b),\, \mathcal{BR}(B \to H b))$ plane (left panel) and in the $(\mathcal{BR}(B \to Z b),\, \mathcal{BR}(B \to A b))$ plane (right panel), for $s^u_R = 0.01$ and $s^d_R = 0.1$.}
\label{fig:doub_lim}
\end{figure*}

In this subsection, we investigate the exclusion behavior of the ($T, B$) doublet scenario within the 2HDM-II, utilizing the parameter-space scan presented in Sec.~\ref{sec:Results}. The anticipated exclusion contours in the correlated planes $(\mathcal{BR}(B \to h b),\ \mathcal{BR}(B \to H b))$ and $(\mathcal{BR}(B \to Z b),\ \mathcal{BR}(B \to A b))$, for the representative choice $s^u_R = 0.01$ and $s^d_R = 0.1$, are displayed in the left and right panels of Fig.~\ref{fig:doub_lim}, respectively.

Remarkably, both $\mathcal{BR}(B \to H b)$ and $\mathcal{BR}(B \to A b)$ can attain values as large as $\sim 88\%$ and 47\%, respectively yielding a total branching fraction into BSM final states of $\mathcal{BR}(B \to \mathrm{BSM}) \simeq 100\%$. As a result, the lower bound on the VLB mass is relaxed from $\sim 1.55$~TeV to approximately $0.98$~TeV. This softening originates from the substantial suppression of the SM-like branching ratios $\mathcal{BR}(B \to h b)$ and $\mathcal{BR}(B \to Z b)$, which are reduced to roughly 6\% each.

The values $s^u_R = 0.01$ and $s^d_R = 0.1$ were chosen deliberately to ensure $\mathcal{BR}(B \to W t) \simeq 0$, consistent with Eq.~\ref{equ:dblt}. This suppression is governed by the right-handed coupling $\propto -s^u_R c^d_R$ (see Table \ref{tab:Wtb})\footnote{The left-handed coupling $V^L_{tB}$ can be safely neglected, as the corresponding mixing angles are highly suppressed: $s^u_L \approx \frac{m_t}{m_T}s^u_R$ and $s^d_L \approx \frac{m_b}{m_B}s_R^d$, as given in Eq.~\ref{ec:rel-angle1}, with $m_T$ and $m_B$ being large and their mass splitting not exceeding 40~GeV~\cite{Arhrib:2024dou, Arhrib:2024tzm}.}, which remains small when $s^u_R \ll s^d_R$. Conversely, if $s^u_R \gtrsim s^d_R$, the $B \to W t$ mode rapidly dominates and can approach 100\%. The same choice $s_R^u \ll s_R^d$ simultaneously enhances the  $T \to Z b$, $T \to h b$, $T \to H b$, and $T \to A b$ couplings. These couplings scale as $s^d_R c^d_R$, as summarized in Tables~\ref{tab:Ztb},~\ref{tab:hBb},~\ref{tab:left_AH} and~\ref{tab:right_AH}.

The same choice of mixing parameters, $s_R^u = 0.01$ and $s_R^d = 0.1$, also leads to
$\mathcal{BR}(B \to H^- t) \simeq 0$. This suppression arises from the structure of
the $B t H^-$ couplings. The left-handed coupling scales as
$\frac{m_t}{m_B}\,\frac{s_R^{u\,2} c_L^d}{s_L^u}$.
Although the smallness of $s_L^u$ tends to enhance this term, the overall
contribution remains negligible due to the strong suppression by the heavy mass
$m_B$ and the small value of $s_R^u = 0.01$.
The right-handed coupling, which is proportional to
$-\frac{s_R^{d\,2} s_L^u}{c_L^d}$, is further suppressed by the tiny value of
$s_L^u$. In addition, the decay $B \to W/H^-T$ is kinematically forbidden and therefore does not contribute, since the mass splitting between the two VLQs satisfies $|m_B - m_T| \lesssim 40$\cite{Arhrib:2024dou, Arhrib:2024tzm}.

\begin{figure}[h!]
\centering
\includegraphics[width=15cm]{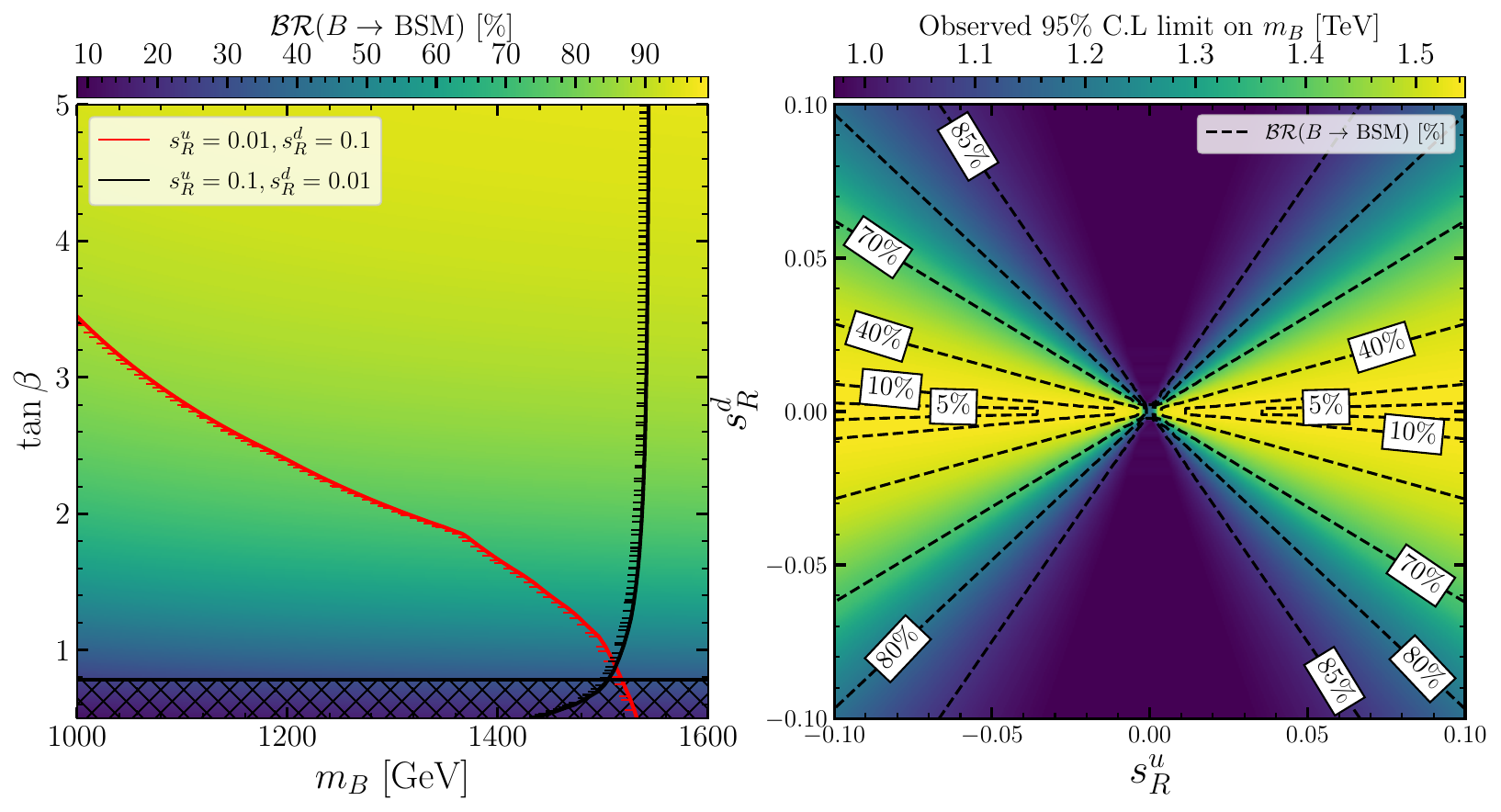}
\caption{The left panel shows exclusion contours in the $(m_B, \tan\beta)$ plane. The region below the red contour is excluded for the configuration $s^u_R = 0.01$ and $s^d_R = 0.1$, while the region to the left of the black contour is excluded for the alternative configuration $s^u_R = 0.1$ and $s^d_R = 0.01$; the branching ratio $\mathcal{BR}(B \to \text{BSM})$ is indicated by the color scale. Additionally, the lower black-shaded area is excluded by the $H^+ \to tb$ search~\cite{ATLAS:2021upq}. The right panel presents the projected 95\% CL exclusion limits on $m_B$ in the $(s^u_R, s^d_R)$ plane for fixed $\tan\beta = 3.5$ and $m_B = 1.5$~TeV, with black dashed contours corresponding to constant values of the branching ratio $\mathcal{BR}(B \to \text{BSM})$. In both panels, all other parameters are fixed to the same values as in Fig.~\ref{fig:exclu_singl}.}
\label{fig:mixing_II}
\end{figure}

To further illustrate the role of the mixing angles, Fig.~\ref{fig:mixing_II} shows the dependence of the VLB mass bounds on these parameters. The left panel presents the excluded regions in the $(m_B, \tan\beta)$ plane for two mixing configurations: (i) $s^u_R = 0.1$, $s^d_R = 0.01$ (black contour) and (ii) $s^u_R = 0.01$, $s^d_R = 0.1$ (red contour). The branching ratio $\mathcal{BR}(B \to \text{BSM})$, calculated for the first configuration, is indicated by the color scale.
For the first configuration, large $m_B$ values are excluded up to approximately 1.54~TeV for $\tan\beta \gtrsim 1$, with a slight relaxation to about 1.52~TeV for $\tan\beta \lesssim 1$. This large exclusion arises from the dominance of the SM decay mode $B \to Wt$, as explained previously. The mild relaxation at low $\tan\beta$ is driven by an increased $\mathcal{BR}(B \to H^- t)$, owing to the left-handed charged-Higgs coupling $Z^L_{Bt}$ dominating over the right-handed coupling $Z^R_{Bt}$ and scaling as $\cot\beta$.
In the inverted configuration, $m_B = 1.54$~TeV is excluded for $\tan\beta \lesssim 1$, but the bound relaxes substantially as $\tan\beta$ increases to reach 1~TeV at $\tan\beta\approx3.45$. This stronger relaxation occurs because $\mathcal{BR}(B \to Hb)$ and $\mathcal{BR}(B \to Ab)$ are proportional to $s^d_R c^d_R \tan\beta$, causing $\mathcal{BR}(B \to \text{BSM})$ to exceed 90\% (as shown by the yellow region in the color scale).

In the right panel, we show the 95\% CL exclusion limit on $m_B$ in the $(s^d_R, s^u_R)$ plane, with the remaining masses fixed as in the previous figures and $\tan\beta = 3.5$.

When $s^d_R > s^u_R$, the exclusion limit on $m_B$ relaxes, reaching approximately 0.98~TeV as the BSM branching fractions increase up to $\gtrsim$~90\% (indicated by the dashed contour). This behaviour occurs for $\tan\beta > 1$ and is reversed for $\tan\beta < 1$.

In contrast, when $s^d_R < s^u_R$, the SM branching fractions dominate---primarily driven by the large $T \to Wb$ branching fraction---resulting in the most stringent exclusion limits, with observed lower bounds $m_B^{\text{obs}} > 1.54$~TeV in the yellow region.

\begin{figure*}[h!]
\centering
\includegraphics[width=15cm]{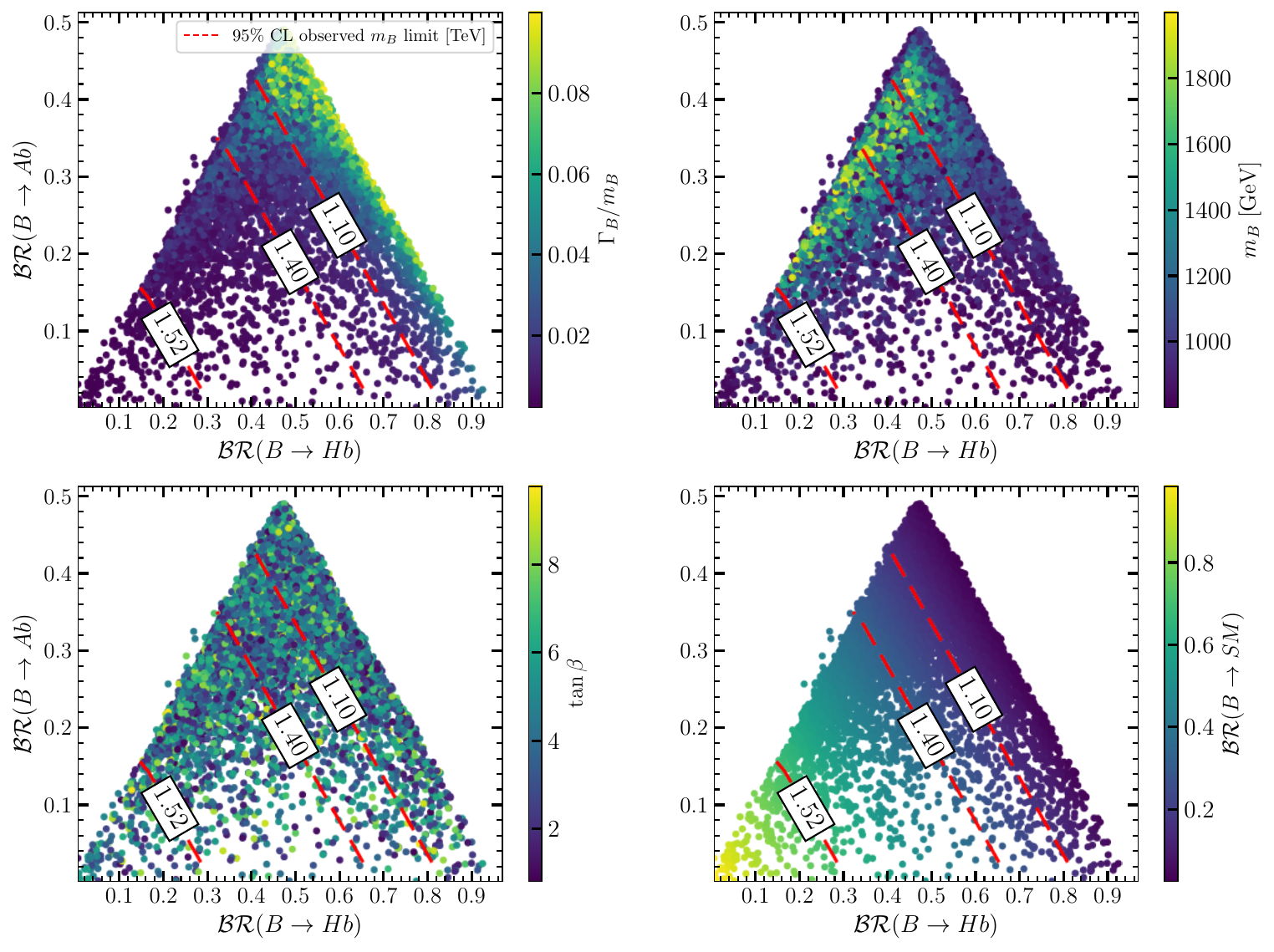}
\caption{Scatter plots of $\mathcal{BR}(B \to Hb)$ versus $\mathcal{BR}(B \to Ab)$ in the VLB doublet scenario (2HDM-II+$TB$), for $s^u_R = 0.01$ and $s^d_R = 0.1$. Colour-coded by $\Gamma_B/m_B$ (top left), $m_B$ (top right), $\tan\beta$ (bottom left), and $\mathcal{BR}(B \to \mathrm{SM})$ (bottom right). Red dashed contours show the observed lower limits on $m_B$.}
\label{fig:doub_params}
\end{figure*}

To assess the dependence of the BSM signatures on key parameters 
in the 2HDM-II+$TB$ scenario, Fig.~\ref{fig:doub_params} displays the BR 
behaviour. The red dashed contours indicate the observed lower bound on $m_B$. 
The BSM BRs $\mathcal{BR}(B \to Hb)$ and $\mathcal{BR}(B \to Ab)$ are shown 
as functions of the total width ratio $\Gamma_B/m_B$, the mass $m_B$, $\tan\beta$, 
and $\mathcal{BR}(B \to \mathrm{SM})$. These BSM modes are enhanced at larger values of 
$\Gamma_B/m_B$ and increase with $m_B$ as well as for intermediate values of $\tan\beta$. 
Conversely, $\mathcal{BR}(B \to \mathrm{SM})$ approaches 100\% when the BSM modes are suppressed. The observed lower limit on $m_B$ weakens as the BRs into BSM decay modes increase.

Table~\ref{tab:BpTB} lists the benchmark points (BPs) for the 2HDM-II+$TB$ scenario. These points are chosen to feature large BSM branching ratios and to remain allowed at 95\% CL, satisfying the condition $\frac{m_B}{m_B^{\mathrm{obs}}} > 1$, where $m_B^{\mathrm{obs}}$ is the observed $m_B$ upper limit at 95\% CL.
\begin{table*}[h!]
\begin{center}
	\setlength{\tabcolsep}{30pt}
	\renewcommand{\arraystretch}{0.8}
	\begin{adjustbox}{max width=\textwidth}
		\begin{tabular}{lccccc}
			\toprule\toprule
			Parameters & BP$_1$ & BP$_2$ & BP$_3$ & BP$_4$ & BP$5$\\\toprule
			\multicolumn{6}{c}{2HDM-II+$TB$ inputs. Masses in GeV.} \\\toprule
			$m_h$ & 125.1 & 125.1 & 125.1 & 125.1 & 125.1\\
			$m_H$ & 559.750 & 580.609 & 492.119 & 614.796 & 596.266\\
			$m_A$ & 585.778 & 730.114 & 576.061 & 777.271 & 686.824\\
			$m{H^\pm}$ & 748.516 & 708.772 & 613.018 & 744.148 & 723.594\\
			$\tan\beta$ & 1.592 & 4.675 & 1.582 & 5.976 & 3.147\\
			$m_B$ & 1075.880 & 1422.386 & 1194.294 & 1342.373 & 1292.381\\
            $m_T$ & 1070.539 & 1415.326 & 1188.366 & 1335.710 & 1285.966\\
			$s^u_R$ & 0.01& 0.01& 0.01 & 0.01 & 0.01\\
			$s^d_R$ & 0.1 & 0.1 & 0.1& 0.1 & 0.1\\
            $s^u_L$ & 0.001& 0.001 & 0.001 & 0.001 & 0.001\\
			$s^d_L$ & 0.0 & 0.0 & 0.0 & 0.0 & 0.0\\
			\toprule
			\multicolumn{6}{c}{BR($B \to XY$) in \%} \\\toprule
			$B \to W^-t$ & 0.128 & 0.183 & 0.145 & 0.168 & 0.133\\
			$B \to Zb$ & 6.915 & 9.627 & 7.773 & 8.887 & 7.059\\
			$B \to hb$ & 6.659 & 9.378 & 7.524 & 8.642 & 6.854 \\
			$B \to Hb$ & 47.788 & 44.847 & 47.392 & 50.817 & 51.744 \\
			$B \to Ab$ & 38.500 & 35.941 & 37.156 & 31.473 & 34.198\\
			$B \to H^-t$ & 0.008 & 0.021 & 0.007 & 0.010 & 0.009\\
			\toprule
			\multicolumn{6}{c}{Total width $\Gamma_B$ [GeV]} \\\toprule
			$\Gamma(B)$ & 29.477 & 48.933 & 35.870 & 44.553 & 50.056\\\toprule
            \multicolumn{6}{c}{Observed $m_B$ limit [GeV]} \\\toprule
            $m_B^{\text{obs}}$ & 1019.263 & 1169.466 & 1066.932 & 1128.647 & 1028.693 \\
			\toprule\toprule
		\end{tabular}
	\end{adjustbox}
\end{center}
\caption{Benchmark points for the 2HDM-II+$TB$ setup.}
\label{tab:BpTB}
\end{table*}
\subsection{2HDM-II with Doublet ($B, Y$)}
\begin{figure*}[h!]
\centering
\includegraphics[width=15cm]{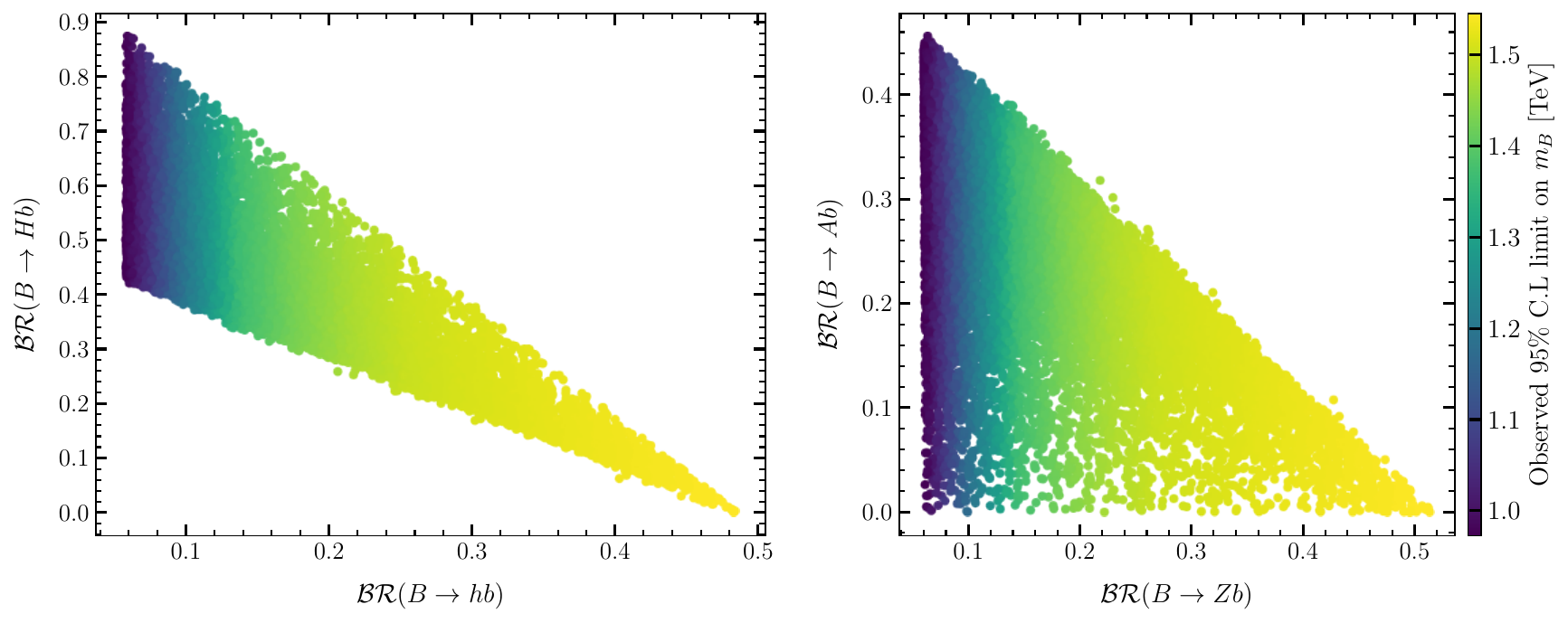}
\caption{Parameter scan results in the 2HDM-II+$BY$ model. The color scale indicates the $m_B$ exclusion limit. Results are shown in the ($\mathcal{BR}(B \to Hb)$, $\mathcal{BR}(B \to hb)$) plane (left) and the ($\mathcal{BR}(B \to Ab)$, $\mathcal{BR}(B \to Zb)$) plane (right).}
\label{fig:obs_by}
\end{figure*}

In this subsection, we investigate the ($B, Y$) doublet scenario within the 2HDM-II framework, in which the coupling to the charged Higgs boson vanishes (see Table \ref{tab:Hp} in Sec.~\ref{sec:cpl}). As a result, Fig.~\ref{fig:obs_by} presents the interplay between BSM\footnote{The $B \to H^+/W^+ Y$ decay is kinematically constrained by the mass splitting being less than 40~GeV~\cite{Arhrib:2024dou}, and is therefore not included among the BSM decay modes.} and SM decay modes by displaying $\mathcal{BR}(B \to Hb)$ versus the SM $\mathcal{BR}(B \to hb)$ (left panel) and $\mathcal{BR}(B \to Ab)$ versus the SM $\mathcal{BR}(B \to hb)$ (right panel). The color bar indicates the observed lower bound on the VLB mass $m_B$, expressed in TeV. The branching fractions $\mathcal{BR}(B \to Hb)$ and $\mathcal{BR}(B \to Ab)$ can reach values as large as $\sim$~88\% and $\sim$~46\%, respectively, which significantly weakens the exclusion power of standard searches and allows the $m_B$ limit to drop below $1~\text{TeV}$. This relaxation is primarily driven by the enhanced coupling strength between the VLB and the neutral Higgs bosons. The remaining SM decay channel, $\mathcal{BR}(B \to Wt)$, is strongly suppressed since its coupling is proportional to $s_L$, which is negligible as implied by Eq.~\ref{ec:rel-angle1}.

\begin{figure*}[h!]
\centering
\includegraphics[width=15cm]{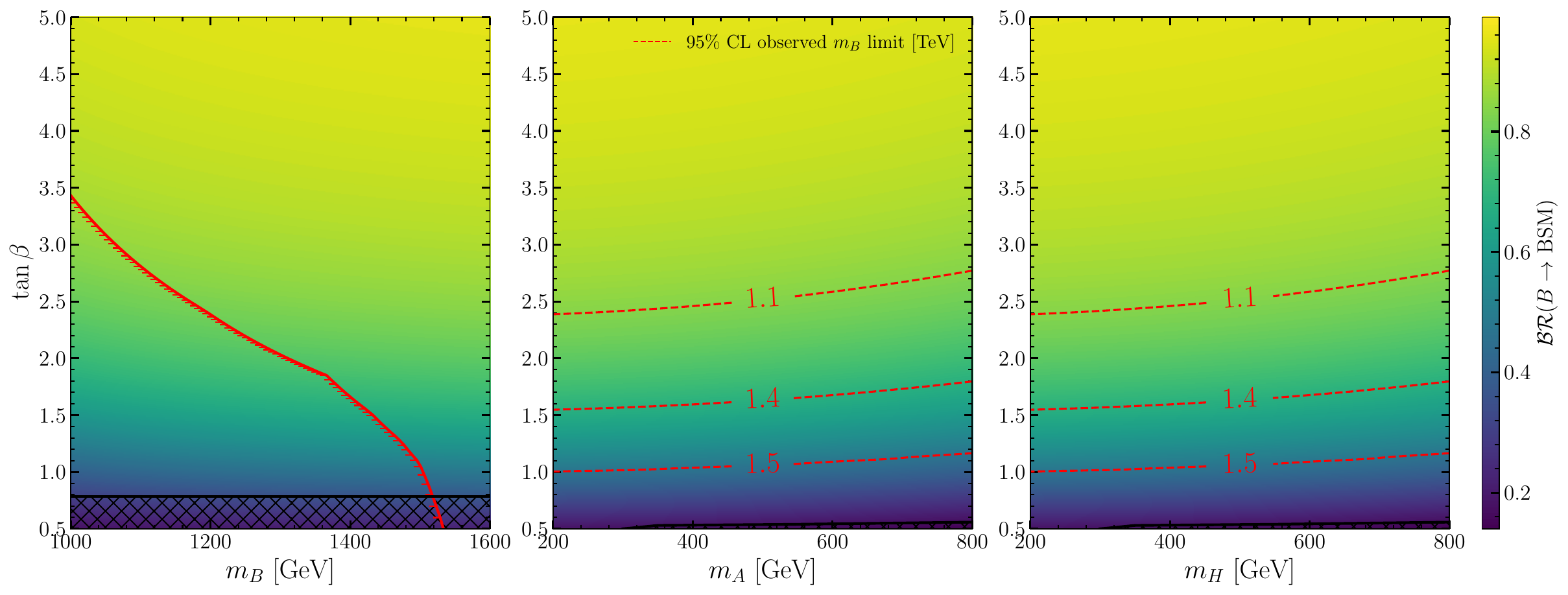}
\caption{$\mathcal{BR}(B \to \text{BSM})$ distributions are shown in the $(m_B, \tan\beta)$ plane (left), the $(m_A, \tan\beta)$ plane (middle), and the $(m_H, \tan\beta)$ plane (right) for the VLB within the 2HDM-II+$BY$ scenario. The red dashed curves in the left panel denote the 95\% CL exclusion, while the lower black-shaded region is excluded by the $H^+ \to tb$ search~\cite{ATLAS:2021upq}. The fixed parameter in the left panel is identical to that in Fig.\ref{fig:exclu_singl}. In the middle and right panels, unranged parameters are fixed to $\{m_B, m_{H^\pm}, s_R\} = \{1.5 \text{ TeV}, 832\text{ GeV}, 0.1\}$ with $m_H=500$~GeV (left panel) and $m_A=500$~GeV (right panel).} 
\label{fig:excl_by}
\end{figure*}

\begin{figure}[H]
\centering
\includegraphics[width=15cm]{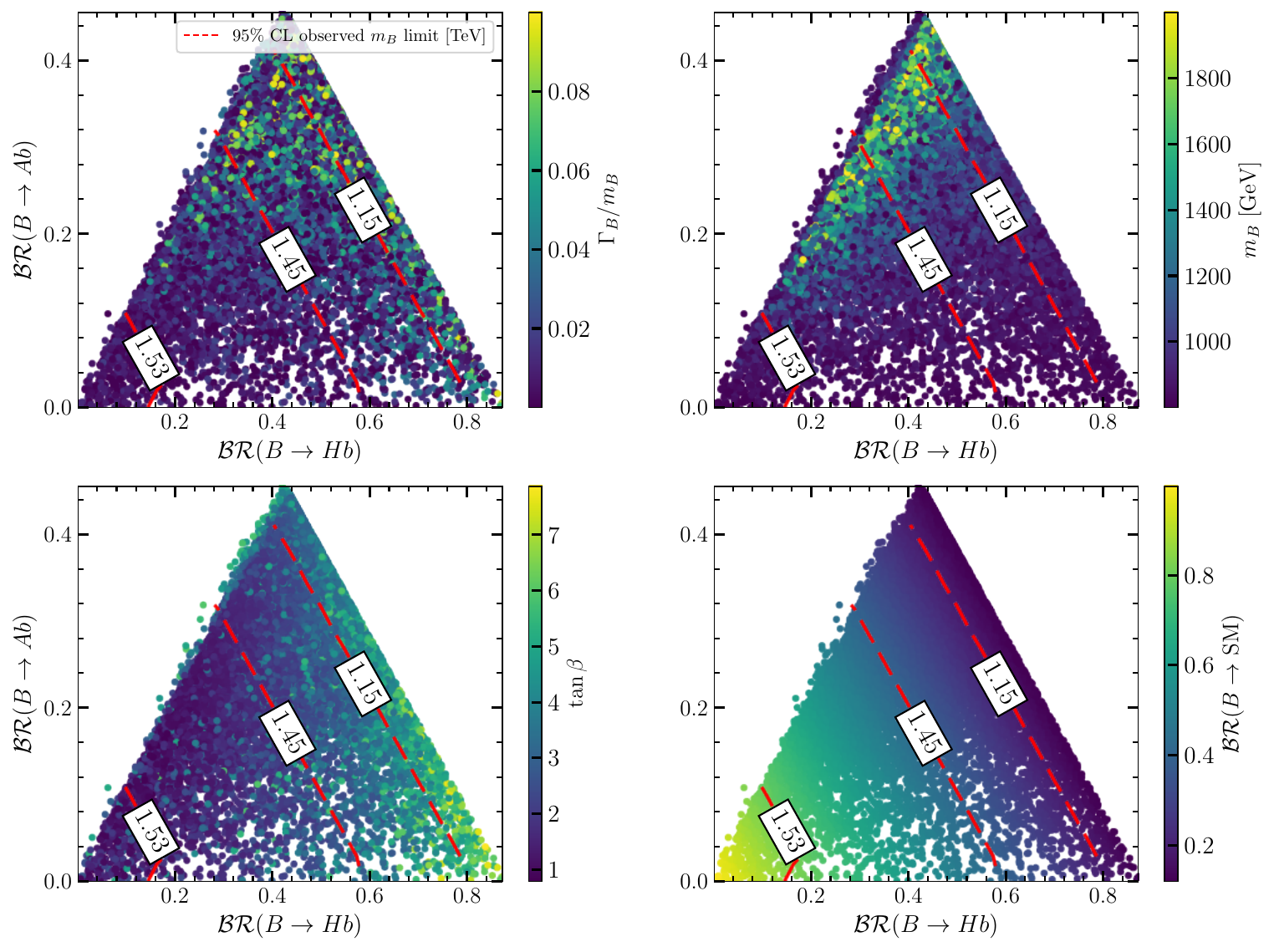}
\caption{Scatter plots of the branching ratios $\mathcal{BR}(B \to Hb)$ versus $\mathcal{BR}(B \to Ab)$ within the 2HDM-II+$BY$ configuration as functions of $\Gamma_B/m_B$ (upper left), $m_B$ (upper right), $\tan\beta$ (lower left), and $\mathcal{BR}(B \to \text{SM})$ (lower right). The red dashed lines indicate the $m_B$ exclusion limit.}
\label{fig:by_params}
\end{figure}

In Fig.~\ref{fig:excl_by}, we present a scan of $\tan\beta$ plotted against $m_B$ (left panel), $m_A$ (middle panel), and $m_H$ (right panel), with colors representing $\mathcal{BR}(B \to \text{BSM})$. The area under the red line shown in the left panel is presenting the recast $m_B$ exclusion region, which weakens significantly for $\tan\beta > 1$, reaching $m_B = 1~\text{TeV}$ at $\tan\beta \sim 3.4$. Beyond $\tan\beta = 3.4$, $\mathcal{BR}(B \to \text{BSM})$ exceeds $\sim$~90\%, rendering $m_B$ unconstrained as SM decay channels become negligible. In the middle and right panels, $m_A$ and $m_H$ show minimal variation, with contour values relaxing only as $\tan\beta$ increases. Thus, the observed $m_B$ relaxation correlates with the rise in $\mathcal{BR}(B \to \text{BSM})$, which scales approximately as $\sim m_B^3\tan^2\beta$.
Fig.~\ref{fig:by_params} displays the branching ratios $\mathcal{BR}(B \to Ab)$ and $\mathcal{BR}(B \to Hb)$ in the 2HDM-II+$BY$ scenario as functions of several key parameters, represented by the color bar: the relative width $\Gamma_B/m_B$ (upper left), the VLB mass $m_B$ (upper right), $\tan\beta$ (lower left), and $\mathcal{BR}(B \to \text{SM})$ (lower right). The relative width increases as $\mathcal{BR}(B \to Ab)$ and $\mathcal{BR}(B \to Hb)$ become more pronounced. The neutral BSM branching ratios remain sizable for both high and low $m_B$ values. Both $\mathcal{BR}(B \to Ab)$ and $\mathcal{BR}(B \to Hb)$ rise with increasing $\tan\beta$, but vanish rapidly for $\tan\beta \lesssim 1$, where $\mathcal{BR}(B \to \text{SM})$ approaches nearly 100\%. In this parameter region, the observed $m_B$ limit is about 1.52~TeV and becomes less stringent as $\mathcal{BR}(B \to \text{SM})$ decreases. These figures clearly demonstrate that the properties of the VLB in the 2HDM+$BY$ model become identical to those in the 2HDM+$TB$ model in the regime where $s^u_R \ll s^d_R$.

Table~\ref{tab:BpTB2} summarizes five benchmark points that satisfy all theoretical and experimental constraints of the 2HDM-II+$BY$ scenario.

\begin{table}[H]
\begin{center}
	\setlength{\tabcolsep}{30pt}
	\renewcommand{\arraystretch}{0.8}
	\begin{adjustbox}{max width=\textwidth}
		\begin{tabular}{lccccc}
			\toprule\toprule
			Parameters & BP$_1$ & BP$_2$ & BP$_3$ & BP$_4$ & BP$5$\\\toprule
			\multicolumn{6}{c}{2HDM-II+$TB$ inputs. Masses in GeV.} \\\toprule
			$m_h$ & 125.1 & 125.1 & 125.1 & 125.1 & 125.1\\ 
			$m_H$ & 595.676 & 646.236 & 642.342 & 584.065 & 498.466\\
			$m_A$ & 746.726 & 678.605 & 733.740 & 776.720 & 533.760\\
			$m{H^\pm}$ & 766.700 & 662.341 & 680.056 & 835.616 & 724.575\\
			$\tan\beta$ & 3.874 & 2.557 & 3.706 & 4.827 & 3.307\\
			$m_B$ & 1149.770 & 1256.946 & 1128.442 & 1013.475 & 1170.200\\
            $m_Y$ & 1147.868 & 1255.956 & 1127.316 & 995.330 & 1170.140\\
			$s_R$ & $-$0.057 & 0.039 & $-$0.044 & 0.188 & $-$0.010\\
			\toprule
			\multicolumn{6}{c}{BR($B \to XY$) in \%} \\\toprule
			$B \to W^-t$ & 0.000 & 0.000 & 0.000 & 0.000 & 0.000\\
			$B \to Zb$ & 6.720 & 11.468 & 7.871 & 6.183 & 6.245\\
			$B \to hb$ & 6.493 & 11.123 & 7.598 & 5.933 & 6.039 \\
			$B \to Hb$ & 53.422 & 40.160 & 48.889 & 63.610 & 45.308\\
			$B \to Ab$ & 33.363 & 37.247 & 35.639 & 24.272 & 42.405\\
			$B \to H^-t$ & 0.0 & 0.0 & 0.0 & 0.0 & 0.0\\
			\toprule
			\multicolumn{6}{c}{Total width $\Gamma_B$ [GeV]} \\\toprule
			$\Gamma(B)$ & 12.323 & 4.497 & 6.008 & 95.281 & 0.435 \\
            \toprule
            \multicolumn{6}{c}{Observed $m_B$ limit [GeV]} \\\toprule
            $m_B^{\text{obs}}$ & 1005.883 & 1248.333 & 1067.651 & 975.834 & 980.447\\
			\toprule\toprule
		\end{tabular}
	\end{adjustbox}
\end{center}
\caption{Benchmark points for the 2HDM-II+$BY$ setup.}
\label{tab:BpTB2}
\end{table}

\section{Conclusion}
\label{sec:Conclusions}

We have studied the collider phenomenology of the 2HDM-II extended by VLB quarks in the singlet ($B$) and doublet (($T, B$), ($B, Y$)) representations. Our analysis shows that when decays into heavy Higgs bosons dominate, the LHC exclusion limits on the VLB mass $m_B$ are substantially weakened. This effect is particularly pronounced in the doublet representations, where BSM branching ratios become large.

In the singlet representation (2HDM-II+$B$), the exclusion bound is reduced to approximately $1.34$~TeV across the full $\tan\beta$ range considered. In contrast, the doublet representations ($B, Y$) and ($T, B$) (in the regime $s^u_R \ll s^d_R$) exhibit significantly weaker limits, reaching about $0.98$~TeV. This reduction is driven by large branching ratios, with $\mathcal{BR}(B \to Hb) \approx 88\%$ and $\mathcal{BR}(B \to Ab) \approx 54\%$ at high $\tan\beta$. From scans over $m_B$ and $\tan\beta$, we find that VLB searches in the 2HDM-II+$TB$ model with $s^u_R = 0.01$ and $s^d_R = 0.1$, as well as in 2HDM-II+$BY$, exclude $m_B \in [1,,1.54]$~TeV at 95\% CL for $\tan\beta \lesssim 3.45$. In the 2HDM-II+$B$ scenario, VLB searches exclude $m_B \lesssim 1.32$~TeV for all $\tan\beta$ values. In both the singlet and doublet scenarios, the exclusion limits show only a mild dependence on the remaining model parameters.

The High-Luminosity LHC will provide a particularly promising opportunity to test this scenario. The large integrated luminosity and improved detector performance should make it possible to directly probe the non-standard decay modes identified here through cascade topologies involving heavy neutral Higgs bosons decaying into $t\bar t$, $b\bar b$, or $\tau^+\tau^-$, as well as charged-Higgs signatures such as $H^- \to tb$ and $H^- \to \tau^-\nu_\tau$, thereby substantially extending the present LHC sensitivity to vector-like bottom quarks in extended Higgs sectors.

\section*{Acknowledgements}
\sloppy
M. Boukidi acknowledges the support of Narodowe Centrum Nauki under OPUS grant no. 2023/49/B/ST2/03862.
\appendix
\section{Light-heavy coupling to SM bosons}
\begin{table}[H]
\centering
\begin{tabular}{c|cccc}
&	$V_{tB}^L$ & $V_{tB}^R$  &	$V_{bY}^L$ & $V_{bY}^R$  \\ \hline
$(B)$	  & $s_L  e^{i\phi}$  & 0 & -& -\\  
$(T, B)$ & $c_L^u s_L^d e^{i\phi_d}- s_L^u c_L^d e^{i\phi_u}$  &$-s_R^u c_R^d e^{i\phi_u}$& -& -\\ 
$(B, Y)$  & $s_L  e^{i\phi}$  &0& $-s_L e^{i\phi}$ &$-s_R e^{i\phi}$  \\
\end{tabular}
\caption{Light-heavy couplings to the $W$ boson.}	
\label{tab:Wtb}
\end{table}
\begin{table}[H]
\centering
\begin{tabular}{c|cc}
& 		$X_{bB}^L$ & $X_{bB}^R$  \\ \hline
$(B)$	& $c_L s_L e^{i\phi}$ &0  \\
$(T, B)$	& 0 & $-s_R^d c_R^d e^{i\phi_d}$ \\
$(B, Y)$	&  $2 c_L s_L e^{i\phi}$ &  $c_R s_R e^{i\phi}$\\
\end{tabular}
\caption{Light-heavy couplings to the $Z$ boson.}	 
\label{tab:Ztb}
\end{table}

\begin{table}[H]
\centering
\begin{tabular}{c|cc}
&  	$Y_{hbB}^L$ & $Y_{hbB}^R$  \\ \hline
$(B)$	& $(s_{\beta\alpha}  -   c_{\beta\alpha} \tan\beta)\frac{m_b}{m_B} c_L s_L e^{i\phi}$ &  $(s_{\beta\alpha}  -    c_{\beta\alpha} \tan\beta) c_L s_L e^{i\phi}$  \\ 
$(T, B)$ & $(s_{\beta\alpha}  -   c_{\beta\alpha} \tan\beta) s_R^d c_R^d e^{i\phi_d}$ &   $(s_{\beta\alpha}  -    c_{\beta\alpha} \tan\beta) \frac{m_b}{m_B} s_R^d c_R^d e^{i\phi_d}$   \\ 
$(B, Y)$ & $(s_{\beta\alpha}  -    c_{\beta\alpha} \tan\beta) c_R s_R e^{i\phi}$ &  $(s_{\beta\alpha}  -   c_{\beta\alpha} \tan\beta)\frac{m_b}{m_B} c_R s_R e^{i\phi}$ \\
\end{tabular}
\caption{Light–heavy left- and right-handed couplings of the SM-like Higgs boson $h$ to the bottom quark.}	
\label{tab:hBb}
\end{table}

\section{Light-heavy coupling to BSM Higgses}
\label{sec:cpl}
\begin{table}[H]
\centering
\begin{tabular}{c|cc}
     & $Y_{HbB}^L$ & $Y_{AbB}^L$ \\ \hline
    $(B)$ & $\left(c_{\beta\alpha} + s_{\beta\alpha} \tan\beta\right) \frac{m_b}{m_B} c_L s_L e^{i\phi}$ & $\tan\beta \frac{m_b}{m_B} c_L s_L e^{i\phi}$ \\ 
    $(T, B)$ & $\left(c_{\beta\alpha} + s_{\beta\alpha} \tan\beta\right) s_R^d c_R^d e^{i\phi_d}$ & $\tan\beta s_R^d c_R^d e^{i\phi_d}$ \\ 
   $(B, Y)$ & $\left(c_{\beta\alpha} + s_{\beta\alpha} \tan\beta\right) c_R s_R e^{i\phi}$ & $\tan\beta c_R s_R e^{i\phi}$ \\ 
\end{tabular}
\caption{Light-heavy left couplings of bottom quarks to the neutral Higgses \{$H, A$\}.}
\label{tab:left_AH}
\end{table}

\begin{table}[H]
\centering
\begin{tabular}{c|cc}
     & $Y_{HbB}^R$ & $Y_{AbB}^R$ \\ \hline
    $(B)$ & $(c_{\beta\alpha} + s_{\beta\alpha} \tan\beta) c_L s_L e^{i\phi}$ & $\tan\beta c_L s_L e^{i\phi}$ \\ 
    $(T, B)$ & $(c_{\beta\alpha} + s_{\beta\alpha} \tan\beta) \frac{m_b}{m_B} s_R^d c_R^d e^{i\phi_d}$ & $\tan\beta \frac{m_b}{m_B} s_R^d c_R^d e^{i\phi_d}$ \\ 
    $(B, Y)$ & $(c_{\beta\alpha} + s_{\beta\alpha} \tan\beta) \frac{m_b}{m_B} c_R s_R e^{i\phi}$ & $\tan\beta \frac{m_b}{m_B} c_R s_R e^{i\phi}$ \\ 
\end{tabular}
\caption{Light-heavy right couplings of bottom quarks to the neutral Higgses \{$H, A$\}.}
\label{tab:right_AH}
\end{table}

\begin{table}[H]
\centering
\begin{tabular}{c|cc}
    & $Z^L_{Bt}$ & $Z^R_{Bt}$ \\ \hline 
    $(B)$ & $s_L$ & 0 \\ 
    $(T, B)$ &  $\frac{m_t}{ m_B}      \left[c_L^u s_L^d e^{ i\phi_d}  +    (  s_R^u{}^2    -  s_L^u{}^2)  \frac{c_L^d}{ s_L^u} e^{i\phi_u} \right]$
    &  $c_L^u s_L^d e^{ i\phi_d}  +  (   s_L^d{}^2 - s_R^d{}^2)  \frac{ s_L^u}{c_L^d} e^{i\phi_u} $   \\  
     $(B, Y)$ & 0 & 0 \\
\end{tabular}
\caption{Heavy-light couplings to the charged Higgs.}
\label{tab:Hp}
\end{table}
\section{VLB decay widths}
\label{sec:dec}
The partial decay widths for the heavy VLB quark are given by the following expressions:

\paragraph{Neutral scalar decay:}
\begin{align}
\Gamma(B \to \phi\, b) &= \frac{g^2}{128\pi}  \frac{m_B}{M_W^2} \lambda^{1/2}(m_B, m_b, M_\phi) 
\nonumber \\
&\times \left\{
(|Y^L_{\phi bB}|^2 + |Y^R_{\phi bB}|^2)\left( 1 + r_b^2 - r_\phi^2 \right) \pm 4 r_b\,  \mathrm{Re} \left( Y^L_{\phi bB} Y^{R^*}_{\phi bB} \right)
\right\},
\label{eq:GammaB}
\end{align}
where the sign $\pm$ corresponds to $+$ for CP-even Higgs (e.g.\ $H$) and $-$ for CP-odd Higgs (e.g.\ $A$). Here, $\phi = H, A$.

\paragraph{Charged scalar decay:}
\begin{align}
\Gamma(B \to H^- t) & = \frac{g^2 }{64 \pi} 
\frac{m_B}{M_W^2} \lambda^{1/2}(m_B,m_t,M_{H^\pm})
\nonumber \\
&\times\left\{ (|Z_{Bt}^L|^2 \cot^2 \beta + |Z_{Bt}^R|^2 \tan^2 \beta )\right. \notag \\
& \left. \times  \left[1+r_t^2 - r_{H^\pm}^2 \right]  + 4 r_t \mathrm{Re}(Z_{Bt}^LZ_{Bt}^{R^*}) \right\} \,.
\label{ec:GammaB}
\end{align}
Here, $r_x = m_x / m_B$, where $x$ refers to one of the decay products, and the function $\lambda(x,y,z)$ is defined as:
\begin{equation}
\lambda(x,y,z) \equiv x^4 + y^4 + z^4 - 2 x^2 y^2 
- 2 x^2 z^2 - 2 y^2 z^2 \,,
\end{equation}

\bibliographystyle{JHEP}
\bibliography{main.bib}

@article{ATLAS:2012yve,
    author = "Aad, Georges and others",
    collaboration = "ATLAS",
    title = "{Observation of a new particle in the search for the Standard Model Higgs boson with the ATLAS detector at the LHC}",
    eprint = "1207.7214",
    archivePrefix = "arXiv",
    primaryClass = "hep-ex",
    reportNumber = "CERN-PH-EP-2012-218",
    doi = "10.1016/j.physletb.2012.08.020",
    journal = "Phys. Lett. B",
    volume = "716",
    pages = "1--29",
    year = "2012"
}

@article{ParticleDataGroup:2020ssz,
    author = "Zyla, P. A. and others",
    collaboration = "Particle Data Group",
    title = "{Review of Particle Physics}",
    doi = "10.1093/ptep/ptaa104",
    journal = "PTEP",
    volume = "2020",
    number = "8",
    pages = "083C01",
    year = "2020"
}

@book{Gunion:1989we,
    author = "Gunion, John F. and Haber, Howard E. and Kane, Gordon L. and Dawson, Sally",
    title = "{The Higgs Hunter's Guide}",
    reportNumber = "SCIPP-89/13, UCD-89-4, BNL-41644",
    doi = "10.1201/9780429496448",
    isbn = "978-0-429-49644-8",
    volume = "80",
    year = "2000"
}

@article{CMS:2012qbp,
    author = "Chatrchyan, Serguei and others",
    collaboration = "CMS",
    title = "{Observation of a New Boson at a Mass of 125 GeV with the CMS Experiment at the LHC}",
    eprint = "1207.7235",
    archivePrefix = "arXiv",
    primaryClass = "hep-ex",
    reportNumber = "CMS-HIG-12-028, CERN-PH-EP-2012-220",
    doi = "10.1016/j.physletb.2012.08.021",
    journal = "Phys. Lett. B",
    volume = "716",
    pages = "30--61",
    year = "2012"
}

@article{Aguilar-Saavedra:2009xmz,
    author = "Aguilar-Saavedra, J. A.",
    title = "{Identifying top partners at LHC}",
    eprint = "0907.3155",
    archivePrefix = "arXiv",
    primaryClass = "hep-ph",
    doi = "10.1088/1126-6708/2009/11/030",
    journal = "JHEP",
    volume = "11",
    pages = "030",
    year = "2009"
}

@article{Okada:2012gy,
    author = "Okada, Yasuhiro and Panizzi, Luca",
    title = "{LHC signatures of vector-like quarks}",
    eprint = "1207.5607",
    archivePrefix = "arXiv",
    primaryClass = "hep-ph",
    reportNumber = "KEK-TH-1560, LYCEN-2012-04",
    doi = "10.1155/2013/364936",
    journal = "Adv. High Energy Phys.",
    volume = "2013",
    pages = "364936",
    year = "2013"
}

@article{Eriksson:2009ws,
    author = "Eriksson, David and Rathsman, Johan and Stal, Oscar",
    title = "{2HDMC: Two-Higgs-Doublet Model Calculator Physics and Manual}",
    eprint = "0902.0851",
    archivePrefix = "arXiv",
    primaryClass = "hep-ph",
    doi = "10.1016/j.cpc.2009.09.011",
    journal = "Comput. Phys. Commun.",
    volume = "181",
    pages = "189--205",
    year = "2010"
}

@article{Bechtle:2020uwn,
    author = "Bechtle, Philip and Heinemeyer, Sven and Klingl, Tobias and Stefaniak, Tim and Weiglein, Georg and Wittbrodt, Jonas",
    title = "{HiggsSignals-2: Probing new physics with precision Higgs measurements in the LHC 13 TeV era}",
    eprint = "2012.09197",
    archivePrefix = "arXiv",
    primaryClass = "hep-ph",
    reportNumber = "BONN-TH-2020-09, DESY-20-228, DESY 20-228, IFT-UAM/CSIC-20-081, LU TP 20-53",
    doi = "10.1140/epjc/s10052-021-08942-y",
    journal = "Eur. Phys. J. C",
    volume = "81",
    number = "2",
    pages = "145",
    year = "2021"
}

@article{Bechtle:2008jh,
    author = "Bechtle, Philip and Brein, Oliver and Heinemeyer, Sven and Weiglein, Georg and Williams, Karina E.",
    title = "{HiggsBounds: Confronting Arbitrary Higgs Sectors with Exclusion Bounds from LEP and the Tevatron}",
    eprint = "0811.4169",
    archivePrefix = "arXiv",
    primaryClass = "hep-ph",
    reportNumber = "DCPT-08-172, IPPP-08-86, BONN-TH-2008-17",
    doi = "10.1016/j.cpc.2009.09.003",
    journal = "Comput. Phys. Commun.",
    volume = "181",
    pages = "138--167",
    year = "2010"
}

@article{Bechtle:2011sb,
    author = "Bechtle, Philip and Brein, Oliver and Heinemeyer, Sven and Weiglein, Georg and Williams, Karina E.",
    title = "{HiggsBounds 2.0.0: Confronting Neutral and Charged Higgs Sector Predictions with Exclusion Bounds from LEP and the Tevatron}",
    eprint = "1102.1898",
    archivePrefix = "arXiv",
    primaryClass = "hep-ph",
    reportNumber = "FR-PHENO-2011-002, BONN-TH-2011-02, DESY-11-016",
    doi = "10.1016/j.cpc.2011.07.015",
    journal = "Comput. Phys. Commun.",
    volume = "182",
    pages = "2605--2631",
    year = "2011"
}

@article{Bechtle:2013wla,
    author = "Bechtle, Philip and Brein, Oliver and Heinemeyer, Sven and St\r{a}l, Oscar and Stefaniak, Tim and Weiglein, Georg and Williams, Karina E.",
    title = "{$\mathsf{HiggsBounds}-4$: Improved Tests of Extended Higgs Sectors against Exclusion Bounds from LEP, the Tevatron and the LHC}",
    eprint = "1311.0055",
    archivePrefix = "arXiv",
    primaryClass = "hep-ph",
    reportNumber = "BONN-TH-2013-21, DESY-13-110",
    doi = "10.1140/epjc/s10052-013-2693-2",
    journal = "Eur. Phys. J. C",
    volume = "74",
    number = "3",
    pages = "2693",
    year = "2014"
}

@article{Bechtle:2015pma,
    author = "Bechtle, Philip and Heinemeyer, Sven and Stal, Oscar and Stefaniak, Tim and Weiglein, Georg",
    title = "{Applying Exclusion Likelihoods from LHC Searches to Extended Higgs Sectors}",
    eprint = "1507.06706",
    archivePrefix = "arXiv",
    primaryClass = "hep-ph",
    reportNumber = "BONN-TH-2015-08, DESY-15-093, SCIPP-15-05",
    doi = "10.1140/epjc/s10052-015-3650-z",
    journal = "Eur. Phys. J. C",
    volume = "75",
    number = "9",
    pages = "421",
    year = "2015"
}

@article{Arhrib:2024dou,
    author = "Arhrib, Abdesslam and Benbrik, Rachid and Boukidi, Mohammed and Moretti, Stefano",
    title = "{Anatomy of vector-like bottom-quark models in the alignment limit of the 2-Higgs doublet model type-II}",
    eprint = "2403.13021",
    archivePrefix = "arXiv",
    primaryClass = "hep-ph",
    doi = "10.1140/epjc/s10052-024-13390-5",
    journal = "Eur. Phys. J. C",
    volume = "84",
    number = "10",
    pages = "1008",
    year = "2024"
}

@article{Arhrib:2024mbq,
    author = "Arhrib, Abdesslam and Benbrik, Rachid and Boukidi, Mohammed and Moretti, Stefano",
    title = "{Large Hadron Collider Signatures of Exotic Vector-Like Quarks within the 2-Higgs Doublet Model Type-II}",
    eprint = "2409.20104",
    archivePrefix = "arXiv",
    primaryClass = "hep-ph",
    month = "9",
    year = "2024"
}

@article{Buchkremer:2013bha,
    author = "Buchkremer, Mathieu and Cacciapaglia, Giacomo and Deandrea, Aldo and Panizzi, Luca",
    title = "{Model Independent Framework for Searches of Top Partners}",
    eprint = "1305.4172",
    archivePrefix = "arXiv",
    primaryClass = "hep-ph",
    reportNumber = "LYCEN-2013-03, SHEP-13-10, CP3-13-22",
    doi = "10.1016/j.nuclphysb.2013.08.010",
    journal = "Nucl. Phys. B",
    volume = "876",
    pages = "376--417",
    year = "2013"
}

@article{Han:2025itd,
    author = "Han, Jin-Zhong and Liu, Yao-Bei and Moretti, Stefano",
    title = "{Searching for single production of vector-like quarks decaying into $Wb$ at a future muon-proton collider}",
    eprint = "2501.01026",
    archivePrefix = "arXiv",
    primaryClass = "hep-ph",
    month = "1",
    year = "2025"
}

@article{Arkani-Hamed:2002iiv,
    author = "Arkani-Hamed, N. and Cohen, A. G. and Katz, E. and Nelson, A. E. and Gregoire, T. and Wacker, Jay G.",
    title = "{The Minimal moose for a little Higgs}",
    eprint = "hep-ph/0206020",
    archivePrefix = "arXiv",
    reportNumber = "BUHEP-02-24, UW-PT-01-09, HUTP-02-A016",
    doi = "10.1088/1126-6708/2002/08/021",
    journal = "JHEP",
    volume = "08",
    pages = "021",
    year = "2002"
}

@article{Han:2003wu,
    author = "Han, Tao and Logan, Heather E. and McElrath, Bob and Wang, Lian-Tao",
    title = "{Phenomenology of the little Higgs model}",
    eprint = "hep-ph/0301040",
    archivePrefix = "arXiv",
    reportNumber = "MADPH-02-1317",
    doi = "10.1103/PhysRevD.67.095004",
    journal = "Phys. Rev. D",
    volume = "67",
    pages = "095004",
    year = "2003"
}

@article{Chang:2003vs,
    author = "Chang, Spencer and He, Hong-Jian",
    title = "{Unitarity of little Higgs models signals new physics of UV completion}",
    eprint = "hep-ph/0311177",
    archivePrefix = "arXiv",
    reportNumber = "HUTP-03-A075, UTHEP-03-19",
    doi = "10.1016/j.physletb.2004.02.027",
    journal = "Phys. Lett. B",
    volume = "586",
    pages = "95--105",
    year = "2004"
}

@article{Yang:2024aav,
    author = "Yang, Bingfang and Li, Zejun and Jia, Xinglong and Moretti, Stefano and Shang, Liangliang",
    title = "{Search for single vector-like B quark production in hadronic final states at the LHC}",
    eprint = "2405.13452",
    archivePrefix = "arXiv",
    primaryClass = "hep-ph",
    doi = "10.1140/epjc/s10052-024-13482-2",
    journal = "Eur. Phys. J. C",
    volume = "84",
    number = "10",
    pages = "1124",
    year = "2024"
}

@article{Contino:2006qr,
    author = "Contino, Roberto and Da Rold, Leandro and Pomarol, Alex",
    title = "{Light custodians in natural composite Higgs models}",
    eprint = "hep-ph/0612048",
    archivePrefix = "arXiv",
    reportNumber = "UAB-FT-619, ROMA1-1445-2006",
    doi = "10.1103/PhysRevD.75.055014",
    journal = "Phys. Rev. D",
    volume = "75",
    pages = "055014",
    year = "2007"
}

@article{Matsedonskyi:2012ym,
    author = "Matsedonskyi, Oleksii and Panico, Giuliano and Wulzer, Andrea",
    title = "{Light Top Partners for a Light Composite Higgs}",
    eprint = "1204.6333",
    archivePrefix = "arXiv",
    primaryClass = "hep-ph",
    doi = "10.1007/JHEP01(2013)164",
    journal = "JHEP",
    volume = "01",
    pages = "164",
    year = "2013"
}

@article{Lodone:2008yy,
    author = "Lodone, Paolo",
    title = "{Vector-like quarks in a 'composite' Higgs model}",
    eprint = "0806.1472",
    archivePrefix = "arXiv",
    primaryClass = "hep-ph",
    doi = "10.1088/1126-6708/2008/12/029",
    journal = "JHEP",
    volume = "12",
    pages = "029",
    year = "2008"
}

@article{Benbrik:2025nfw,
    author = "Benbrik, R. and Berrouj, M. and Boukidi, M. and Kahime, K.",
    title = "{Exploring vector-like top quark pair production via charged Higgs decays in multi-b-jet and opposite-sign dilepton final state at the LHC}",
    doi = "10.1140/epjc/s10052-025-14237-3",
    journal = "Eur. Phys. J. C",
    volume = "85",
    number = "5",
    pages = "500",
    year = "2025"
}

@article{Martin:2009iq,
    author = "Martin, A. D. and Stirling, W. J. and Thorne, R. S. and Watt, G.",
    title = "{Parton distributions for the LHC}",
    eprint = "0901.0002",
    archivePrefix = "arXiv",
    primaryClass = "hep-ph",
    reportNumber = "IPPP-08-95, DCPT-08-190, CAVENDISH-HEP-08-16",
    doi = "10.1140/epjc/s10052-009-1072-5",
    journal = "Eur. Phys. J. C",
    volume = "63",
    pages = "189--285",
    year = "2009"
}

@article{Martin:2009bu,
    author = "Martin, A. D. and Stirling, W. J. and Thorne, R. S. and Watt, G.",
    title = "{Uncertainties on alpha(S) in global PDF analyses and implications for predicted hadronic cross sections}",
    eprint = "0905.3531",
    archivePrefix = "arXiv",
    primaryClass = "hep-ph",
    reportNumber = "IPPP-09-33, DCPT-09-66, CAVENDISH-HEP-09-06",
    doi = "10.1140/epjc/s10052-009-1164-2",
    journal = "Eur. Phys. J. C",
    volume = "64",
    pages = "653--680",
    year = "2009"
}

@article{Martin:2010db,
    author = "Martin, A. D. and Stirling, W. J. and Thorne, R. S. and Watt, G.",
    title = "{Heavy-quark mass dependence in global PDF analyses and 3- and 4-flavour parton distributions}",
    eprint = "1007.2624",
    archivePrefix = "arXiv",
    primaryClass = "hep-ph",
    reportNumber = "IPPP-10-29, DCPT-10-58, CAVENDISH-HEP-10-07, CERN-PH-TH-2010-160",
    doi = "10.1140/epjc/s10052-010-1462-8",
    journal = "Eur. Phys. J. C",
    volume = "70",
    pages = "51--72",
    year = "2010"
}

@article{Benbrik:2022kpo,
    author = "Benbrik, Rachid and Boukidi, Mohammed and Moretti, Stefano",
    title = "{Probing charged Higgs bosons in the two-Higgs-doublet model type II with vectorlike quarks}",
    eprint = "2211.07259",
    archivePrefix = "arXiv",
    primaryClass = "hep-ph",
    doi = "10.1103/PhysRevD.109.055016",
    journal = "Phys. Rev. D",
    volume = "109",
    number = "5",
    pages = "055016",
    year = "2024"
}

@article{Abouabid:2023mbu,
    author = "Abouabid, Hamza and Arhrib, Abdesslam and Benbrik, Rachid and Boukidi, Mohammed and Falaki, Jaouad El",
    title = "{The oblique parameters in the 2HDM with vector-like quarks: confronting M $_{W}$ CDF-II anomaly}",
    eprint = "2302.07149",
    archivePrefix = "arXiv",
    primaryClass = "hep-ph",
    doi = "10.1088/1361-6471/ad3f34",
    journal = "J. Phys. G",
    volume = "51",
    number = "7",
    pages = "075001",
    year = "2024"
}

@article{Benbrik:2024hsf,
    author = "Benbrik, Rachid and Berrouj, Mbark and Boukidi, Mohammed",
    title = "{Investigation of charged Higgs bosons production from vectorlike T quark decays at e{\ensuremath{\gamma}} collider}",
    eprint = "2408.15985",
    archivePrefix = "arXiv",
    primaryClass = "hep-ph",
    doi = "10.1103/PhysRevD.111.015027",
    journal = "Phys. Rev. D",
    volume = "111",
    number = "1",
    pages = "015027",
    year = "2025"
}

@article{Arhrib:2024nbj,
    author = "Arhrib, Abdesslam and Benbrik, Rachid and Berrouj, Mbark and Boukidi, Mohammed and Manaut, Bouzid",
    title = "{Search for charged Higgs bosons through vectorlike top quark pair production at the LHC}",
    eprint = "2407.01348",
    archivePrefix = "arXiv",
    primaryClass = "hep-ph",
    doi = "10.1103/PhysRevD.111.095026",
    journal = "Phys. Rev. D",
    volume = "111",
    number = "9",
    pages = "095026",
    year = "2025"
}

@article{Benbrik:2024bxt,
    author = "Benbrik, Rachid and Boukidi, Mohammed and Moretti, Stefano",
    title = "{Charged Higgs Boson Mass Bounds in 2HDM-II: Impact of Vector-Like Quarks}",
    eprint = "2409.16054",
    archivePrefix = "arXiv",
    primaryClass = "hep-ph",
    doi = "10.22323/1.476.0083",
    journal = "PoS",
    volume = "ICHEP2024",
    pages = "083",
    year = "2025"
}

@article{Dermisek:2019vkc,
    author = "Derm{\'\i}{\v{s}}ek, Radovan and Lunghi, Enrico and Shin, Seodong",
    title = "{Hunting for Vectorlike Quarks}",
    eprint = "1901.03709",
    archivePrefix = "arXiv",
    primaryClass = "hep-ph",
    reportNumber = "EFI-19-1",
    doi = "10.1007/JHEP04(2019)019",
    journal = "JHEP",
    volume = "04",
    pages = "019",
    year = "2019",
    note = "[Erratum: JHEP 10, 058 (2020)]"
}

@article{Ghosh:2023xhs,
    author = "Ghosh, Anupam and Konar, Partha",
    title = "{Precision prediction of a democratic up-family philic KSVZ axion model at the LHC}",
    eprint = "2305.08662",
    archivePrefix = "arXiv",
    primaryClass = "hep-ph",
    doi = "10.1016/j.dark.2024.101746",
    journal = "Phys. Dark Univ.",
    volume = "47",
    pages = "101746",
    year = "2025"
}

@article{Benbrik:2019zdp,
    author = "Benbrik, Rachid and others",
    title = "{Signatures of vector-like top partners decaying into new neutral scalar or pseudoscalar bosons}",
    eprint = "1907.05929",
    archivePrefix = "arXiv",
    primaryClass = "hep-ph",
    doi = "10.1007/JHEP05(2020)028",
    journal = "JHEP",
    volume = "05",
    pages = "028",
    year = "2020"
}

@article{Gunion:1992hs,
    author = "Gunion, John F. and Haber, Howard E. and Kane, Gordon L. and Dawson, Sally",
    title = "{Errata for the Higgs hunter's guide}",
    eprint = "hep-ph/9302272",
    archivePrefix = "arXiv",
    reportNumber = "SCIPP-92-58",
    month = "12",
    year = "1992"
}

@article{Branco:2011iw,
    author = "Branco, G. C. and Ferreira, P. M. and Lavoura, L. and Rebelo, M. N. and Sher, Marc and Silva, Joao P.",
    title = "{Theory and phenomenology of two-Higgs-doublet models}",
    eprint = "1106.0034",
    archivePrefix = "arXiv",
    primaryClass = "hep-ph",
    doi = "10.1016/j.physrep.2012.02.002",
    journal = "Phys. Rept.",
    volume = "516",
    pages = "1--102",
    year = "2012"
}

@article{Agashe:2004rs,
    author = "Agashe, Kaustubh and Contino, Roberto and Pomarol, Alex",
    title = "{The Minimal composite Higgs model}",
    eprint = "hep-ph/0412089",
    archivePrefix = "arXiv",
    reportNumber = "UAB-FT-567",
    doi = "10.1016/j.nuclphysb.2005.04.035",
    journal = "Nucl. Phys. B",
    volume = "719",
    pages = "165--187",
    year = "2005"
}

@article{Bellazzini:2014yua,
    author = "Bellazzini, Brando and Cs\'aki, Csaba and Serra, Javi",
    title = "{Composite Higgses}",
    eprint = "1401.2457",
    archivePrefix = "arXiv",
    primaryClass = "hep-ph",
    doi = "10.1140/epjc/s10052-014-2766-x",
    journal = "Eur. Phys. J. C",
    volume = "74",
    number = "5",
    pages = "2766",
    year = "2014"
}

@article{Arhrib:2024tzm,
    author = "Arhrib, Abdesslam and Benbrik, Rachid and Boukidi, Mohammed and Manaut, Bouzid and Moretti, Stefano",
    title = "{Anatomy of vector-like top-quark models in the alignment limit of the 2-Higgs Doublet Model Type-II}",
    eprint = "2401.16219",
    archivePrefix = "arXiv",
    primaryClass = "hep-ph",
    doi = "10.1140/epjc/s10052-024-13692-8",
    journal = "Eur. Phys. J. C",
    volume = "85",
    number = "1",
    pages = "2",
    year = "2025"
}

@article{Hewett:1988xc,
    author = "Hewett, JoAnne L. and Rizzo, Thomas G.",
    title = "{Low-Energy Phenomenology of Superstring Inspired E(6) Models}",
    reportNumber = "MAD-PH-446, IS-J-3005",
    doi = "10.1016/0370-1573(89)90071-9",
    journal = "Phys. Rept.",
    volume = "183",
    pages = "193",
    year = "1989"
}

@article{ATLAS:2024gyc,
    author = "Aad, Georges and others",
    collaboration = "ATLAS",
    title = "{Search for pair-production of vector-like quarks in lepton+jets final states containing at least one b-tagged jet using the Run 2 data from the ATLAS experiment}",
    eprint = "2401.17165",
    archivePrefix = "arXiv",
    primaryClass = "hep-ex",
    reportNumber = "CERN-EP-2023-254",
    doi = "10.1016/j.physletb.2024.138743",
    journal = "Phys. Lett. B",
    volume = "854",
    pages = "138743",
    year = "2024"
}

@article{He:1999vp,
    author = "He, Hong-Jian and Tait, Timothy M. P. and Yuan, C. P.",
    title = "{New top flavor models with seesaw mechanism}",
    eprint = "hep-ph/9911266",
    archivePrefix = "arXiv",
    reportNumber = "MSUHEP-91015, ANL-HEP-PR-99-115",
    doi = "10.1103/PhysRevD.62.011702",
    journal = "Phys. Rev. D",
    volume = "62",
    pages = "011702",
    year = "2000"
}

@article{Wang:2013jwa,
    author = "Wang, Xu-Feng and Du, Chun and He, Hong-Jian",
    title = "{LHC Higgs Signatures from Topflavor Seesaw Mechanism}",
    eprint = "1304.2257",
    archivePrefix = "arXiv",
    primaryClass = "hep-ph",
    doi = "10.1016/j.physletb.2013.05.015",
    journal = "Phys. Lett. B",
    volume = "723",
    pages = "314--323",
    year = "2013"
}

@article{ATLAS:2022tla,
    author = "Aad, Georges and others",
    collaboration = "ATLAS",
    title = "{Search for pair-produced vector-like top and bottom partners in events with large missing transverse momentum in pp collisions with the ATLAS detector}",
    eprint = "2212.05263",
    archivePrefix = "arXiv",
    primaryClass = "hep-ex",
    reportNumber = "CERN-EP-2022-201",
    doi = "10.1140/epjc/s10052-023-11790-7",
    journal = "Eur. Phys. J. C",
    volume = "83",
    number = "8",
    pages = "719",
    year = "2023"
}

@article{CMS:2018dcw,
    author = "Sirunyan, Albert M and others",
    collaboration = "CMS",
    title = "{Search for single production of vector-like quarks decaying to a top quark and a W boson in proton-proton collisions at $\sqrt{s} =$ 13 TeV}",
    eprint = "1809.08597",
    archivePrefix = "arXiv",
    primaryClass = "hep-ex",
    reportNumber = "CMS-B2G-17-018, CERN-EP-2018-230",
    doi = "10.1140/epjc/s10052-019-6556-3",
    journal = "Eur. Phys. J. C",
    volume = "79",
    pages = "90",
    year = "2019"
}

@article{Bechtle:2020pkv,
    author = "Bechtle, Philip and Dercks, Daniel and Heinemeyer, Sven and Klingl, Tobias and Stefaniak, Tim and Weiglein, Georg and Wittbrodt, Jonas",
    title = "{HiggsBounds-5: Testing Higgs Sectors in the LHC 13 TeV Era}",
    eprint = "2006.06007",
    archivePrefix = "arXiv",
    primaryClass = "hep-ph",
    reportNumber = "BONN-TH-2020-03, DESY 20-093, DESY-20-093, IFT-UAM/CSIC-20-072, LU 20-27",
    doi = "10.1140/epjc/s10052-020-08557-9",
    journal = "Eur. Phys. J. C",
    volume = "80",
    number = "12",
    pages = "1211",
    year = "2020"
}

@article{Benbrik:2023xlo,
    author = "Benbrik, R. and Berrouj, M. and Boukidi, M. and Habjia, A. and Ghourmin, E. and Rahili, L.",
    title = "{Search for single production of vector-like top partner T\(\rightarrow\)H$^+$b and H\(^\pm\)\(\rightarrow\)tb\textasciimacron{} at the LHC Run-III}",
    doi = "10.1016/j.physletb.2023.138024",
    journal = "Phys. Lett. B",
    volume = "843",
    pages = "138024",
    year = "2023"
}

@article{Bahl:2022igd,
    author = {Bahl, Henning and Biek\"otter, Thomas and Heinemeyer, Sven and Li, Cheng and Paasch, Steven and Weiglein, Georg and Wittbrodt, Jonas},
    title = "{HiggsTools: BSM scalar phenomenology with new versions of HiggsBounds and HiggsSignals}",
    eprint = "2210.09332",
    archivePrefix = "arXiv",
    primaryClass = "hep-ph",
    doi = "10.1016/j.cpc.2023.108803",
    journal = "Comput. Phys. Commun.",
    volume = "291",
    pages = "108803",
    year = "2023"
}

@article{Benbrik:2024fku,
    author = "Benbrik, Rachid and Boukidi, Mohammed and Ech-chaouy, Mohamed and Moretti, Stefano and Salime, Khawla and Yan, Qi-Shu",
    title = "{Vector-Like Quarks at the LHC: A unified perspective from ATLAS and CMS exclusion limits}",
    eprint = "2412.01761",
    archivePrefix = "arXiv",
    primaryClass = "hep-ph",
    doi = "10.1007/JHEP03(2025)020",
    journal = "JHEP",
    volume = "03",
    pages = "020",
    year = "2025"
}

@article{Bhardwaj:2022nko,
    author = "Bhardwaj, Akanksha and Mandal, Tanumoy and Mitra, Subhadip and Neeraj, Cyrin",
    title = "{Roadmap to explore vectorlike quarks decaying to a new scalar or pseudoscalar}",
    eprint = "2203.13753",
    archivePrefix = "arXiv",
    primaryClass = "hep-ph",
    doi = "10.1103/PhysRevD.106.095014",
    journal = "Phys. Rev. D",
    volume = "106",
    number = "9",
    pages = "095014",
    year = "2022"
}

@article{Kanemura:1993hm,
    author = "Kanemura, Shinya and Kubota, Takahiro and Takasugi, Eiichi",
    title = "{Lee-Quigg-Thacker bounds for Higgs boson masses in a two doublet model}",
    eprint = "hep-ph/9303263",
    archivePrefix = "arXiv",
    reportNumber = "OS-GE-32-93",
    doi = "10.1016/0370-2693(93)91205-2",
    journal = "Phys. Lett. B",
    volume = "313",
    pages = "155--160",
    year = "1993"
}

@article{ATLAS:2021upq,
    author = "Aad, Georges and others",
    collaboration = "ATLAS",
    title = "{Search for charged Higgs bosons decaying into a top quark and a bottom quark at $ \sqrt{\mathrm{s}} $ = 13 TeV with the ATLAS detector}",
    eprint = "2102.10076",
    archivePrefix = "arXiv",
    primaryClass = "hep-ex",
    reportNumber = "CERN-EP-2021-004",
    doi = "10.1007/JHEP06(2021)145",
    journal = "JHEP",
    volume = "06",
    pages = "145",
    year = "2021"
}

@article{ATLAS:2018gfm,
    author = "Aaboud, Morad and others",
    collaboration = "ATLAS",
    title = "{Search for charged Higgs bosons decaying via $H^{\pm} \to \tau^{\pm}\nu_{\tau}$ in the $\tau$+jets and $\tau$+lepton final states with 36 fb$^{-1}$ of $pp$ collision data recorded at $\sqrt{s} = 13$ TeV with the ATLAS experiment}",
    eprint = "1807.07915",
    archivePrefix = "arXiv",
    primaryClass = "hep-ex",
    reportNumber = "CERN-EP-2018-148",
    doi = "10.1007/JHEP09(2018)139",
    journal = "JHEP",
    volume = "09",
    pages = "139",
    year = "2018"
}

@article{ATLAS:2020gxx,
    author = "Aad, Georges and others",
    collaboration = "ATLAS",
    title = "{Search for a heavy Higgs boson decaying into a Z boson and another heavy Higgs boson in the $\ell \ell bb$ and $\ell \ell WW$ final states in $pp$ collisions at $\sqrt{s}=13$ $\text {TeV}$ with the ATLAS detector}",
    eprint = "2011.05639",
    archivePrefix = "arXiv",
    primaryClass = "hep-ex",
    reportNumber = "CERN-EP-2020-191",
    doi = "10.1140/epjc/s10052-021-09117-5",
    journal = "Eur. Phys. J. C",
    volume = "81",
    number = "5",
    pages = "396",
    year = "2021"
}

@article{Arhrib:2016rlj,
    author = "Arhrib, A. and Benbrik, R. and King, S. J. D. and Manaut, B. and Moretti, S. and Un, C. S.",
    title = "{Phenomenology of 2HDM with vectorlike quarks}",
    eprint = "1607.08517",
    archivePrefix = "arXiv",
    primaryClass = "hep-ph",
    doi = "10.1103/PhysRevD.97.095015",
    journal = "Phys. Rev. D",
    volume = "97",
    pages = "095015",
    year = "2018"
}

@article{Barroso:2013awa,
    author = "Barroso, A. and Ferreira, P. M. and Ivanov, I. P. and Santos, Rui",
    title = "{Metastability bounds on the two Higgs doublet model}",
    eprint = "1303.5098",
    archivePrefix = "arXiv",
    primaryClass = "hep-ph",
    doi = "10.1007/JHEP06(2013)045",
    journal = "JHEP",
    volume = "06",
    pages = "045",
    year = "2013"
}

@article{Deshpande:1977rw,
    author = "Deshpande, Nilendra G. and Ma, Ernest",
    title = "{Pattern of Symmetry Breaking with Two Higgs Doublets}",
    reportNumber = "OITS-81",
    doi = "10.1103/PhysRevD.18.2574",
    journal = "Phys. Rev. D",
    volume = "18",
    pages = "2574",
    year = "1978"
}

@article{ATLAS:2020zms,
    author = "Aad, Georges and others",
    collaboration = "ATLAS",
    title = "{Search for heavy Higgs bosons decaying into two tau leptons with the ATLAS detector using $pp$ collisions at $\sqrt{s}=13$ TeV}",
    eprint = "2002.12223",
    archivePrefix = "arXiv",
    primaryClass = "hep-ex",
    reportNumber = "CERN-EP-2020-014",
    doi = "10.1103/PhysRevLett.125.051801",
    journal = "Phys. Rev. Lett.",
    volume = "125",
    number = "5",
    pages = "051801",
    year = "2020"
}

@article{Gopalakrishna:2015wwa,
    author = "Gopalakrishna, Shrihari and Mukherjee, Tuhin Subhra and Sadhukhan, Soumya",
    title = "{Extra neutral scalars with vectorlike fermions at the LHC}",
    eprint = "1504.01074",
    archivePrefix = "arXiv",
    primaryClass = "hep-ph",
    doi = "10.1103/PhysRevD.93.055004",
    journal = "Phys. Rev. D",
    volume = "93",
    number = "5",
    pages = "055004",
    year = "2016"
}

@article{Chang:1999nh,
    author = "Chang, Sanghyeon and Hisano, Junji and Nakano, Hiroaki and Okada, Nobuchika and Yamaguchi, Masahiro",
    title = "{Bulk standard model in the Randall-Sundrum background}",
    eprint = "hep-ph/9912498",
    archivePrefix = "arXiv",
    reportNumber = "TU-581, KEK-TH-665, NIIG-DP-99-3",
    doi = "10.1103/PhysRevD.62.084025",
    journal = "Phys. Rev. D",
    volume = "62",
    pages = "084025",
    year = "2000"
}

@article{Gherghetta:2000qt,
    author = "Gherghetta, Tony and Pomarol, Alex",
    title = "{Bulk fields and supersymmetry in a slice of AdS}",
    eprint = "hep-ph/0003129",
    archivePrefix = "arXiv",
    reportNumber = "CERN-TH-2000-081, UNIL-IPT-00-06",
    doi = "10.1016/S0550-3213(00)00392-8",
    journal = "Nucl. Phys. B",
    volume = "586",
    pages = "141--162",
    year = "2000"
}

@article{Contino:2003ve,
    author = "Contino, Roberto and Nomura, Yasunori and Pomarol, Alex",
    title = "{Higgs as a Holographic Pseudo Goldstone Boson}",
    eprint = "hep-ph/0306259",
    archivePrefix = "arXiv",
    reportNumber = "FT-UAM-03-11, FERMILAB-PUB-03-195-T, UAB-FT-549",
    doi = "10.1016/j.nuclphysb.2003.08.027",
    journal = "Nucl. Phys. B",
    volume = "671",
    pages = "148--174",
    year = "2003"
}

@article{Schmaltz:2002wx,
    author = "Schmaltz, Martin",
    editor = "Bentvelsen, S. and de Jong, P. and Koch, J. and Laenen, Eric",
    title = "{Physics beyond the standard model (theory): Introducing the little Higgs}",
    eprint = "hep-ph/0210415",
    archivePrefix = "arXiv",
    reportNumber = "BUHEP-02-35, BUPUB-02-35",
    doi = "10.1016/S0920-5632(03)01409-9",
    journal = "Nucl. Phys. B Proc. Suppl.",
    volume = "117",
    pages = "40--49",
    year = "2003"
}

@article{Grimus:2007if,
    author = "Grimus, W. and Lavoura, L. and Ogreid, O. M. and Osland, P.",
    title = "{A Precision constraint on multi-Higgs-doublet models}",
    eprint = "0711.4022",
    archivePrefix = "arXiv",
    primaryClass = "hep-ph",
    reportNumber = "UWTHPH-2007-28",
    doi = "10.1088/0954-3899/35/7/075001",
    journal = "J. Phys. G",
    volume = "35",
    pages = "075001",
    year = "2008"
}

@article{Angelescu:2015uiz,
    author = "Angelescu, Andrei and Djouadi, Abdelhak and Moreau, Gr{\'e}gory",
    title = "{Scenarii for interpretations of the LHC diphoton excess: two Higgs doublets and vector-like quarks and leptons}",
    eprint = "1512.04921",
    archivePrefix = "arXiv",
    primaryClass = "hep-ph",
    reportNumber = "LPT-ORSAY-15-99",
    doi = "10.1016/j.physletb.2016.02.064",
    journal = "Phys. Lett. B",
    volume = "756",
    pages = "126--132",
    year = "2016"
}

@article{Aguilar-Saavedra:2002phh,
    author = "Aguilar-Saavedra, J. A.",
    title = "{Effects of mixing with quark singlets}",
    eprint = "hep-ph/0210112",
    archivePrefix = "arXiv",
    reportNumber = "FISIST-16-2002-CFIF",
    doi = "10.1103/PhysRevD.69.099901",
    journal = "Phys. Rev. D",
    volume = "67",
    pages = "035003",
    year = "2003",
    note = "[Erratum: Phys.Rev.D 69, 099901 (2004)]"
}

@article{Benbrik:2025kvz,
    author = "Benbrik, R. and Berrouj, M. and Boukidi, M. and Ech-chaouy, M. and Kahime, K. and Salime, K.",
    title = "{Relaxing vector-like top quark mass limits through exotic decays in the type-II two-Higgs-doublet model}",
    doi = "10.1140/epjc/s10052-025-15047-3",
    journal = "Eur. Phys. J. C",
    volume = "85",
    number = "11",
    pages = "1275",
    year = "2025"
}

@article{CMS:2025zwi,
    author = "Hayrapetyan, Aram and others",
    collaboration = "CMS",
    title = "{Search for single production of a vector-like T quark decaying to a top quark and a neutral scalar boson in the lepton+jets final state in proton-proton collisions at $\sqrt{s}$ = 13 TeV}",
    eprint = "2510.25874",
    archivePrefix = "arXiv",
    primaryClass = "hep-ex",
    reportNumber = "CMS-B2G-23-009, CERN-EP-2025-228",
    month = "10",
    year = "2025"
}

@article{Han:2022zgw,
    author = "Han, Lin and Shen, Jie-Fen and Liu, Yao-Bei",
    title = "{Searching for a vector-like $B$ quark through a $tW$ decay channel at future electron{\textendash}positron colliders}",
    doi = "10.1140/epjc/s10052-022-10527-2",
    journal = "Eur. Phys. J. C",
    volume = "82",
    number = "7",
    pages = "637",
    year = "2022"
}

@article{Yang:2025ktj,
    author = "Yang, Shuo and Wang, Yi-Hang and Zhao, Peng-Bo and Ma, Ji-Long",
    title = "{Search for heavy vector-like B quark via pair production in fully hadronic channels at CLIC*}",
    eprint = "2504.15882",
    archivePrefix = "arXiv",
    primaryClass = "hep-ph",
    doi = "10.1088/1674-1137/adf183",
    journal = "Chin. Phys.",
    volume = "49",
    number = "11",
    pages = "113101",
    year = "2025"
}

@article{CMS:2024bni,
    author = "Hayrapetyan, Aram and others",
    collaboration = "CMS",
    title = "{Review of searches for vector-like quarks, vector-like leptons, and heavy neutral leptons in proton{\textendash}proton collisions at {\ensuremath{\sqrt{}}}s=13 TeV at the CMS experiment}",
    eprint = "2405.17605",
    archivePrefix = "arXiv",
    primaryClass = "hep-ex",
    reportNumber = "CMS-EXO-23-006, CERN-EP-2024-095",
    doi = "10.1016/j.physrep.2024.09.012",
    journal = "Phys. Rept.",
    volume = "1115",
    pages = "570--677",
    year = "2025"
}

@article{He:2001fz,
    author = "He, Hong-Jian and Hill, Christopher T. and Tait, Timothy M. P.",
    title = "{Top Quark Seesaw, Vacuum Structure and Electroweak Precision Constraints}",
    eprint = "hep-ph/0108041",
    archivePrefix = "arXiv",
    reportNumber = "UTEXAS-HEP-01-013, FERMILAB-PUB-01-164-T, ANL-HEP-PR-01-047",
    doi = "10.1103/PhysRevD.65.055006",
    journal = "Phys. Rev. D",
    volume = "65",
    pages = "055006",
    year = "2002"
}

@article{He:2014ora,
    author = "He, Hong-Jian and Xianyu, Zhong-Zhi",
    title = "{Extending Higgs Inflation with TeV Scale New Physics}",
    eprint = "1405.7331",
    archivePrefix = "arXiv",
    primaryClass = "hep-ph",
    doi = "10.1088/1475-7516/2014/10/019",
    journal = "JCAP",
    volume = "10",
    pages = "019",
    year = "2014"
}

@article{CMS:2022fck,
    author = "Tumasyan, Armen and others",
    collaboration = "CMS",
    title = "{Search for pair production of vector-like quarks in leptonic final states in proton-proton collisions at $ \sqrt{s} $ = 13 TeV}",
    eprint = "2209.07327",
    archivePrefix = "arXiv",
    primaryClass = "hep-ex",
    reportNumber = "CMS-B2G-20-011, CERN-EP-2022-175",
    doi = "10.1007/JHEP07(2023)020",
    journal = "JHEP",
    volume = "07",
    pages = "020",
    year = "2023"
}

@article{ATLAS:2021uiz,
    author = "Aad, Georges and others",
    collaboration = "ATLAS",
    title = "{Search for resonances decaying into photon pairs in 139 fb$^{-1}$ of $pp$ collisions at $\sqrt {s}$=13 TeV with the ATLAS detector}",
    eprint = "2102.13405",
    archivePrefix = "arXiv",
    primaryClass = "hep-ex",
    reportNumber = "CERN-EP-2020-248",
    doi = "10.1016/j.physletb.2021.136651",
    journal = "Phys. Lett. B",
    volume = "822",
    pages = "136651",
    year = "2021"
}

@article{CMS:2022goy,
    author = "Tumasyan, Armen and others",
    collaboration = "CMS",
    title = "{Searches for additional Higgs bosons and for vector leptoquarks in $\tau\tau$ final states in proton-proton collisions at $\sqrt{s}$ = 13 TeV}",
    eprint = "2208.02717",
    archivePrefix = "arXiv",
    primaryClass = "hep-ex",
    reportNumber = "CMS-HIG-21-001, CERN-EP-2022-137",
    doi = "10.1007/JHEP07(2023)073",
    journal = "JHEP",
    volume = "07",
    pages = "073",
    year = "2023"
}

@article{CMS:2019pzc,
    author = "Sirunyan, Albert M and others",
    collaboration = "CMS",
    title = "{Search for heavy Higgs bosons decaying to a top quark pair in proton-proton collisions at $\sqrt{s} =$ 13 TeV}",
    eprint = "1908.01115",
    archivePrefix = "arXiv",
    primaryClass = "hep-ex",
    reportNumber = "CMS-HIG-17-027, CERN-EP-2019-147",
    doi = "10.1007/JHEP04(2020)171",
    journal = "JHEP",
    volume = "04",
    pages = "171",
    year = "2020",
    note = "[Erratum: JHEP 03, 187 (2022)]"
}

@article{CMS:2020imj,
    author = "Sirunyan, Albert M and others",
    collaboration = "CMS",
    title = "{Search for charged Higgs bosons decaying into a top and a bottom quark in the all-jet final state of pp collisions at $ \sqrt{s} $ = 13 TeV}",
    eprint = "2001.07763",
    archivePrefix = "arXiv",
    primaryClass = "hep-ex",
    reportNumber = "CMS-HIG-18-015, CERN-EP-2019-277",
    doi = "10.1007/JHEP07(2020)126",
    journal = "JHEP",
    volume = "07",
    pages = "126",
    year = "2020"
}

@article{CMS:2019bfg,
    author = "Sirunyan, Albert M and others",
    collaboration = "CMS",
    title = "{Search for charged Higgs bosons in the H$^{\pm}$ $\to$ $\tau^{\pm}\nu_\tau$ decay channel in proton-proton collisions at $\sqrt{s} =$ 13 TeV}",
    eprint = "1903.04560",
    archivePrefix = "arXiv",
    primaryClass = "hep-ex",
    reportNumber = "CMS-HIG-18-014, CERN-EP-2019-025",
    doi = "10.1007/JHEP07(2019)142",
    journal = "JHEP",
    volume = "07",
    pages = "142",
    year = "2019"
}

@article{ATLAS:2021vrm,
    collaboration = "ATLAS",
    title = "{Combined measurements of Higgs boson production and decay using up to $139$ fb$^{-1}$ of proton-proton collision data at $\sqrt{s}= 13$ TeV collected with the ATLAS experiment}",
    reportNumber = "ATLAS-CONF-2021-053",
    year = "2021"
}

@article{CMS:2021mku,
    author = "Tumasyan, A. and others",
    collaboration = "CMS",
    title = "{Search for a heavy resonance decaying into a top quark and a W boson in the lepton+jets final state at $ \sqrt{s} $ = 13 TeV}",
    eprint = "2111.10216",
    archivePrefix = "arXiv",
    primaryClass = "hep-ex",
    reportNumber = "CMS-B2G-20-010, CERN-EP-2021-223",
    doi = "10.1007/JHEP04(2022)048",
    journal = "JHEP",
    volume = "04",
    pages = "048",
    year = "2022"
}

@article{Cingiloglu:2023ylm,
    author = "Cingiloglu, Kivanc Y. and Frank, Mariana",
    title = "{Vacuum stability and electroweak precision in the two-Higgs-doublet model with vectorlike quarks}",
    eprint = "2309.03700",
    archivePrefix = "arXiv",
    primaryClass = "hep-ph",
    doi = "10.1103/PhysRevD.109.036016",
    journal = "Phys. Rev. D",
    volume = "109",
    number = "3",
    pages = "036016",
    year = "2024"
}

@article{Han:2022jcp,
    author = "Han, Jin-Zhong and Liu, Yao-Bei and Xing, Lu and Xu, Shuai",
    title = "{Search for single production of vectorlike B-quarks at the LHC*}",
    eprint = "2208.06845",
    archivePrefix = "arXiv",
    primaryClass = "hep-ph",
    doi = "10.1088/1674-1137/ac79ab",
    journal = "Chin. Phys. C",
    volume = "46",
    number = "10",
    pages = "103103",
    year = "2022"
}

@article{Benbrik:2026zjv,
    author = "Benbrik, R. and Berrouj, M. and Boukidi, M. and Chatoui, H. and Ech-chaouy, M. and Kahime, K. and Salime, K.",
    title = "{Single Production of a Vector-Like Top as a Probe of Charged Higgs Bosons at a Muon-Proton Collider}",
    eprint = "2601.07758",
    archivePrefix = "arXiv",
    primaryClass = "hep-ph",
    month = "1",
    year = "2026"
}

@article{Benbrik:2023quz,
    author = "Benbrik, Rachid and Boukidi, Mohammed",
    title = "{Phenomenology of Heavy Quark at the LHC}",
    doi = "10.5772/intechopen.1001607",
    month = "12",
    year = "2023"
}
\end{document}